\documentclass[11pt]{article}
\usepackage{amssymb,amsmath,amsfonts}
\usepackage{graphicx}
\usepackage{graphics}
\usepackage{eepic,epsfig}

\@addtoreset{equation}{section}

\makeatother

\begin{document}

\title{Finite temperature charge and current densities around a cosmic
string in AdS spacetime with compact dimension}
\author{E. R. Bezerra de Mello$^{1}$ \thanks{%
E-mail: emello@fisica.ufpb.br},\thinspace\ W. Oliveira dos Santos$^{1}$%
\thanks{%
E-mail: wagner.physics@gmail.com}, \thinspace\ A. A. Saharian$^{2}$\thanks{%
E-mail: saharian@ysu.am} \\
\\
$^{1}$\textit{Departamento de F\'{\i}sica, Universidade Federal da Para\'{\i}%
ba}\\
\textit{58.059-970, Caixa Postal 5.008, Jo\~{a}o Pessoa, PB, Brazil}\vspace{%
0.3cm}\\
$^{2}$\textit{Department of Physics, Yerevan State University,}\\
\textit{1 Alex Manoogian Street, 0025 Yerevan, Armenia}}
\maketitle

\begin{abstract}
For a massive scalar field with a general curvature coupling parameter, we
investigate the finite temperature contributions to the Hadamard function
and to the charge and current densities in the geometry of a magnetic flux
carrying generalized cosmic string embedded in $(D+1)$-dimensional locally
AdS spacetime with a compactified spatial dimension. For $D=4$, the geometry
on the AdS boundary, in the context of the AdS/CFT duality, corresponds to a
cosmic string as a linear defect, compactified along its axis. In contrast
to the case of the Minkowski bulk, the upper bound on the chemical potential
does not depend on the field mass and is completely determined by the length
of compact dimension and by the enclosed magnetic flux. The only nonzero
components correspond to the charge density, to the azimuthal current and to
the current along the compact dimension. They are periodic functions of
magnetic fluxes with the period equal to the flux quantum. The charge
density is an odd function and the currents are even functions of the
chemical potential. At high temperatures the influence of the gravitational
field and topology on the charge density is subdominant and the leading term
in the corresponding expansion coincides with that for the charge density in
the Minkowski spacetime. The current densities are topology-induced
quantities and their behavior at high temperatures is completely different
with the linear dependence on the temperature. At small temperatures and for
chemical potentials smaller than the critical value, the thermal expectation
values are exponentially suppressed for both massive and massless fields.
\end{abstract}

\section{Introduction}

According to the Big Bang theory, in the early stages the Universe was
hotter and in a more symmetric state. During the expansion it has cooled
down and underwent series of phase transitions accompanied by spontaneous
breakdown of symmetries which could result in the formation of various types
of topological defects \cite{Kibble,V-S}. They include domain walls, cosmic
strings and monopoles. Among these defects the cosmic strings are of special
interest. They are linear structures of trapped energy density, analogous to
defects such as vortex lines in superconductors and superfluids. The
influence of cosmic strings on the geometry of surrounding space can be of
cosmological and astrophysical significance in a large number of phenomena,
such as producing cosmic microwave background anisotropies, non-Gaussianity
and B-mode polarization, sourcing gravitational waves and high energy cosmic
rays, gravitational lensing of astrophysical objects \cite{Hindmarsh2}. The
parameter that characterizes the strength of gravitational interactions of
strings with matter is its tension, that is given in natural units by $G\mu
_{0}$, being $G$ the Newton's constant and $\mu _{0}$ is the linear mass
density, proportional to the square of the symmetry breaking energy scale.
Another mechanism for the formation of cosmic string type defects has been
considered recently in brane inflationary models (for reviews see \cite%
{Cope11,Cher15}).

In the simplest model, the gravitational field produced by a cosmic string
is approximated by a planar angle deficit in the two-dimensional subspace
orthogonal to the string. Although the corresponding local geometry outside
the cosmic string core is flat, the nontrivial topology is a source of
interesting effects in quantum field theory. In particular, the vacuum
expectation values of physical quantities bilinear in the field operator are
shifted by an amount that depends on the planar angle deficit. Among those
quantities, the vacuum energy-momentum tensor is of special interest as an
important local characteristic and also as the source for the gravitational
back-reaction of quantum effects. The vacuum polarization for different spin
fields has been widely investigated in the literature (see, for example
references given in \cite{Bell14}). As an additional physical
characteristic, the cosmic string may carry a magnetic flux along its axis
and this is another source for topological influence on the properties of
the vacuum, being an effect of Aharonov-Bohm-type. Among the physical
manifestations we mention here the appearance of vacuum currents circulating
around the cosmic string \cite{Srir01,Site09}. The ground state currents in
a (2+1)-dimensional conical spacetime with applications to graphene
nanocones (described in terms of the effective Dirac model) have been
investigated in \cite{Site08}-\cite{Bell20d3}.

All these investigations have been done for conical defects embedded in flat
spacetime. For both scenarios of cosmic string formation mentioned above,
the background geometry is not flat and it is of interest to study the
influence of the gravitational field on the cosmic string induced effects on
the properties of quantum vacuum. For cosmic strings formed in the early
Universe during the inflationary era the spacetime geometry is well
approximated by de Sitter spacetime and the polarization of the Bunch-Davies
vacuum for scalar, fermionic and electromagnetic fields by a cosmic string
in that geometry is investigated in references \cite%
{Beze09,Beze10dS,Saha17dS}. The background geometry in the most part of
braneworld models is described by another maximally symmetric solution of
the Einstein field equations, namely, by anti-de Sitter (AdS) spacetime. An
additional motivation for investigations of quantum field-theoretical
effects in AdS bulk comes from the AdS/CFT correspondence that provides a
duality relation between two-different theories: string theory or
supergravity on the AdS background and conformal field theory localized on
the AdS boundary. The local properties of the scalar and fermionic vacua
around a cosmic string in AdS spacetime were studied in \cite%
{Beze12AdS,Beze13AdS}.

Another source of topological quantum effects is the compactification of
spatial dimensions. The compact dimensions are an inherent feature of most
high-energy theories in fundamental physics. They also appear in effective
field-theoretical descriptions of a number of condensed matter systems like
graphene nanoutubes and nanoloops (see, e.g., \cite{Cast09}). The
compactification gives rise to the Casimir type contributions in the
expectation values of physical quantities that depend on the
compactification length and on the periodicity conditions along respective
dimensions. The effects of the compactification on the local properties of
the quantum vacuum in problems with cosmic strings on background of flat
spacetime have been considered in references \cite{Beze12comp}-\cite%
{Brag19comp}. The combined effects of cosmic string, compactification and of
the gravitational field were discussed in \cite{Oliv19}-\cite{Bell22b} for
the background AdS geometry.

It is of interest to note that the finite temperature effects can also be
interpreted as a special type of compactification along the Euclidean time
coordinate with the period equal to the inverse temperature. This important
prescription is used to fix the relation between the vacuum and finite
temperature two-point functions in quantum field theory. In the present
paper we investigate the finite temperature effects on the expectation
values of the charge and current densities for a scalar field in background
of AdS spacetime in the presence of a cosmic string type defect and a
compactified dimension. Thermal Green functions and\ the finite temperature
expectation value of the energy-momentum tensor for scalar field around a
cosmic string embedded in (3+1)-dimensional Minkowski spacetime have been
considered in \cite{Davi88}-\cite{More17}. The finite temperature charge and
current densities for a scalar field in the geometry of a compactified
cosmic string are investigated in \cite{Moha16}. The thermal effects in AdS
spacetime may have qualitatively new features compared with the flat
spacetime background. A well-known example comes from the thermodynamics of
black-holes. As it has been shown in \cite{Hawk83}, the black holes in AdS
spacetime have a minimum temperature that corresponds to the horizon radius
of the order of AdS curvature scale. An interesting topic of investigations
in the context of the AdS/CFT correspondence is the duality between the
theories in the bulk and on the AdS boundary at finite temperature (see, for
instance, \cite{Elia19} and references given therein). In particular, the
thermal field theory on the boundary is dual to a bulk theory with an AdS
black hole having the same temperature \cite{Witt98}. The two-point function
and the expectation values of the field squared and energy-momentum tensor
for a conformally invariant scalar field at finite temperature on background
of AdS spacetime were studied in \cite{Alle87}. The results of recent
investigations of the finite temperature effects for scalar and fermionic
fields are presented in \cite{Ambr17,Ambr18} (for thermal effects in
braneworld see, for example, \cite{Brev01,Cham01,Chat17} and references
therein).

This paper is organized as follows. In the next section the background
geometry, the field content and the complete set of scalar modes are
presented. These modes are used for the evaluation of the thermal Hadamard
function. In Section \ref{Current}, by using the decomposed Hadamard
function, general expressions are derived for thermal contributions to the
charge density and to the azimuthal and axial current densities. Various
special cases of the general results are considered in Section \ref%
{sec:Special}. The analysis of the expectation values in different
asymptotic regions for the values of the parameters is presented in Section %
\ref{sec:Asymp}. Section \ref{sec:Conc} summarizes the main results obtained
in the paper.

\section{Background geometry and the thermal Hadamard function}

\label{sec2}

The background geometry we are going to consider is described by the
following $(D+1)$-dimensional line element: 
\begin{equation}
ds^{2}=e^{-2y/a}\left[ dt^{2}-dr^{2}-r^{2}d\phi ^{2}-\left( dx^{3}\right)
^{2}-d\mathbf{x}_{\parallel }^{2}\right] -dy^{2}\ ,  \label{ds1}
\end{equation}%
where $a$ is a constant and the Cartesian coordinates $\mathbf{x}_{\parallel
}=(x^{4},\ldots ,x^{D-1})$ cover the $(D-4)$-dimensional subspace. For the
variations ranges of the coordinates one has $-\infty <t,x^{l},y<+\infty $, $%
l=4,\ldots ,D-1$, $r\geq 0$ and $\phi \in \lbrack 0,\ 2\pi /q]$. It will be
assumed that the coordinate $x^{3}=z$ is compactified to a circle with
length $L$ and, hence, $0\leq z\leq L$. For the parameter $q$ we take $q\geq
1$. In the special case $q=1$ and for decompactified coordinate $z$, $%
-\infty <z<+\infty $, the geometry described by (\ref{ds1}) presents the AdS
spacetime covered by cylindrical Poincar\'{e} coordinates and sourced by the
negative cosmological constant $\Lambda =-D(D-1)a^{-2}$. The corresponding
Ricci scalar is expressed as $R=-D(D+1)a^{-2}$. For $q>1$ and $r>0$ the
local geometry is the same as that for $(D+1)$-dimensional AdS spacetime. In
this case the curvature tensor has delta function type singularity located
on the $(D-2)$-dimensional hypersurface $r=0$ and described by the line
element 
\begin{equation}
ds_{\mathrm{c}}^{2}=e^{-2y/a}(dt^{2}-dz^{2}-d\mathbf{x}_{\parallel
}^{2})-dy^{2}.  \label{dsc}
\end{equation}%
The latter corresponds to a $(D-1)$-dimensional AdS spacetime. In the
special case $D=3$ it describes a linear defect that corresponds to an
idealized model of cosmic string in AdS spacetime with the linear mass
density $\mu _{0}$. For this particular example the parameter $q$ is
expressed in terms of the linear mass density $\mu _{0}$ by the relation $%
q^{-1}=1-4G\mu _{0}$ with $G$ being the gravitational constant. For $-\infty
<z<+\infty $ the topology of the $(D-2)$-dimensional hypersurface $r=0$ is
trivial. The compactification of the coordinate $z$ does not change the
local geometry and the topology of the core becomes cylindrical, $%
R^{D-3}\times S^{1}$. Similar to that case of AdS spacetime, introducing the
coordinate $w=ae^{y/a}$, with the variation range $w\in \lbrack 0,\ \infty )$%
, the line element is rewritten as 
\begin{equation}
ds^{2}=\left( \frac{a}{w}\right) ^{2}\left( dt^{2}-dr^{2}-r^{2}d\phi
^{2}-dz^{2}-d\mathbf{x}_{\parallel }^{2}-dw^{2}\right) \ .  \label{ds2}
\end{equation}%
This shows the conformal relation of the problem under consideration with
the corresponding problem on the Minkowski bulk with compactified coordinate 
$z$. Note that for the special case $D=4$ the part $d\mathbf{x}_{\parallel
}^{2}$ is absent in (\ref{ds2}) and the background geometry of the conformal
field on the AdS boundary, in the context of the AdS/CFT duality,
corresponds to the standard cosmic string as a linear defect compactified
along its axis.

We are interested in the effects of finite temperature and compactification
on the expectation values of the current density for a charged massive
scalar field $\varphi (x)$ in background of the geometry described above.
Assuming the non-minimal coupling to the curvature with the parameter $\xi $
and in the presence of a classical gauge field $A_{l}$, the field equation
reads%
\begin{equation}
(g^{kl}D_{k}D_{l}+m^{2}+\xi R)\varphi (x)=0\ ,  \label{KGE}
\end{equation}%
being $D_{l}=\nabla _{l}+ieA_{l}$ the gauge extended covariant derivative
operator. For minimal and conformal couplings one has $\xi =0$ and $\xi
=(D-1)/(4D)$, respectively. In the discussion below it will be convenient to
work in the coordinates corresponding to (\ref{ds2}). The coordinate $z$ is
compact and in addition to the field equation one needs to specify the
periodicity condition along the corresponding direction. We will impose the
condition 
\begin{equation}
\varphi (t,r,\phi ,z+L,\mathbf{x}_{\parallel },w)=e^{2\pi i\delta }\varphi
(t,r,\phi ,z,\mathbf{x}_{\parallel },w),  \label{QPC}
\end{equation}%
where $\delta $ is a constant parameter. As to the classical vector
potential $A_{l}$, a simple configuration $A_{l}=(0,0,A_{2},A_{3},0,...,0)$
with constant covariant components $A_{2}$ and $A_{3}$ will be taken. Of
course, we could take nonzero constant components along noncompact
coordinates, but they are removed from the problem by a linear gauge
transformation. Similar transformations for the components $A_{2}$ and $%
A_{3} $ change the phases of the periodicity conditions along the respective
directions and those components are physically relevant. Their effects are
topological. We can express the components of the vector potential in terms
of the corresponding magnetic fluxes $\Phi _{2}$ and $\Phi _{3}$ as $%
A_{2}=-q\Phi _{2}/(2\pi )$ and $A_{3}=-\Phi _{3}/L$.

In the problem under consideration the properties of a given state for
quantum scalar field are obtained from the corresponding two point
functions. Here we will use the Hadamard function. Assuming that the field
is prepared in an equilibrium state with temperature $T$, it is defined as
the expectation value 
\begin{equation}
G(x,x^{\prime })=\mathrm{tr}\left[ \hat{\rho}\left( \varphi (x)\varphi
^{\dagger }(x^{\prime })+\varphi ^{\dagger }(x^{\prime })\varphi (x)\right) %
\right] \ ,  \label{C}
\end{equation}%
with $\hat{\rho}=Z^{-1}e^{-\beta (\hat{H}-\mu ^{\prime }\hat{Q})}$ being the
density matrix. Here, $\beta =1/T$, $\hat{H}$ is the Hamiltonian operator, $%
\hat{Q}$ is a conserved charge with the corresponding chemical potential $%
\mu ^{\prime }$. As usual, $Z=\mathrm{tr}[e^{-\beta (\hat{H}-\mu ^{\prime }%
\hat{Q})}]$ is the grand-canonical partition function. By expanding the
field operator in terms of the complete set of positive and negative energy
mode functions $\varphi _{\sigma }^{(\pm )}(x)$, with the energies $\pm
E_{\sigma }$, and using the properties of the annihilation and creation
operators, the Hadamard function is presented in the form (the details are
similar to the procedure used in \cite{Beze13T} for the problem in Minkowski
spacetime with toroidally compact dimensions)%
\begin{equation}
G(x,x^{\prime })=G_{0}(x,x^{\prime })+G_{T}(x,x^{\prime })\ ,
\label{W_total}
\end{equation}%
where $G_{0}(x,x^{\prime })=\sum_{\mathbf{\sigma }}\sum_{u=\pm }\varphi
_{\sigma }^{(u)}(x)\varphi _{\sigma }^{(u)\ast }(x^{\prime })$ is the zero
temperature Hadamard function and%
\begin{equation}
G_{T}(x,x^{\prime })=2\sum_{\sigma }\sum_{u=\pm }\frac{\varphi _{\sigma
}^{(u)}(x)\varphi _{\sigma }^{(u)\ast }(x^{\prime })}{e^{\beta (E_{\sigma
}-u\mu )}-1},  \label{WF1}
\end{equation}%
is the thermal part. Here, $\mu =e\mu ^{\prime }$ and the set of quantum
numbers $\sigma $ specifies the modes. The symbol $\sum_{\sigma }$ stands
for summation over discrete quantum numbers and integration over the
continuum ones. In (\ref{WF1}), the parts with $u=+$ and $u=-$ correspond to
the contributions of the particles and antiparticles. The chemical
potentials for them have opposite signs.

The normalized mode functions for the geometry at hand, obeying the
periodicity condition (\ref{QPC}), are given in \cite{Oliv19}. They are
expressed as 
\begin{equation}
\varphi _{\sigma }^{(\pm )}(x)=\sqrt{\frac{qa^{1-D}\lambda p}{2(2\pi
)^{D-3}LE_{\sigma }}}w^{\frac{D}{2}}J_{\nu }(pw)J_{q|n+\alpha |}(\lambda
r)e^{iqn\phi +ik_{l}z+i\mathbf{k}\cdot \mathbf{x}_{\parallel }\mp iE_{\sigma
}t}\ ,  \label{COS}
\end{equation}%
with $(p,\lambda )\in \lbrack 0,\ \infty )$, $n=0,\pm 1,\ \pm 2,\ ...$, and $%
\mathbf{k}=(k^{4},\ldots ,k^{D-1})$ corresponds to the components of the
momentum in the subspace with the coordinates $\mathbf{x}_{\parallel }$. The
order of the Bessel function $J_{\nu }(z)$ is defined by 
\begin{equation}
\nu =\sqrt{\frac{{D^{2}}}{4}+a^{2}m^{2}-\xi D(D+1)}\ ,  \label{nu}
\end{equation}%
and in the order of the Bessel function for the radial part we use the
notation $\alpha =eA_{2}/q=-\Phi _{2}/\Phi _{0}$, being $\Phi _{0}=2\pi /e$
the flux quantum. The eigenvalues for the component $k^{3}$ of the momentum
along the compact dimension $z$ are discretized by the condition (\ref{QPC}%
): 
\begin{equation}
k^{3}=k_{l}=\frac{2\pi }{L}(l+\delta )\ ,\;l=0,\pm 1,\ \pm 2,\ ...
\label{k3}
\end{equation}%
The energy of the modes is given by 
\begin{equation}
E_{\sigma }=\sqrt{\lambda ^{2}+p^{2}+\mathbf{k}^{2}+\tilde{k}_{l}^{2}},
\label{Energy}
\end{equation}%
where 
\begin{equation}
\tilde{k}_{l}=2\pi \frac{l+\tilde{\delta}}{L},\;\tilde{\delta}=\delta +\frac{%
eA_{3}L}{2\pi }=\delta -\frac{\Phi _{3}}{\Phi _{0}}.  \label{Phase}
\end{equation}%
In this way the set of quantum numbers $\sigma $ is specified to $\sigma
=(p,\lambda ,n,l,\mathbf{k})$ and the collective summation is understood as%
\begin{equation}
\sum_{\sigma }=\int d\mathbf{k}\int_{0}^{\infty }dp\int_{0}^{\infty
}d\lambda \ \sum_{l,n=-\infty }^{+\infty }\ .  \label{Sumsig}
\end{equation}

The function $G_{0}(x,x^{\prime })$ is obtained from the corresponding
Wightman function given in \cite{Oliv19}: 
\begin{eqnarray}
G_{0}(x,x^{\prime }) &=&\frac{q(ww^{\prime })^{\frac{D}{2}}}{(2\pi
)^{D-3}a^{D-1}L}\sum_{\sigma }\frac{p\lambda }{E_{\sigma }}e^{inq\Delta \phi
+ik_{l}\Delta z+i\mathbf{k}\cdot \Delta \mathbf{x}_{\parallel }}  \notag \\
&&\times \cos \left( E_{\sigma }\Delta t\right) J_{q|n+\alpha |}(\lambda
r)J_{q|n+\alpha |}(\lambda r^{\prime })J_{\nu }(pw)J_{\nu }(pw^{\prime }),
\label{wight_0}
\end{eqnarray}%
where $\Delta t=t-t^{\prime }$, $\Delta \phi =\phi -\phi ^{\prime }$, $%
\Delta z=z-z^{\prime }$, $\Delta \mathbf{x}_{\parallel }=\mathbf{x}%
_{\parallel }-\mathbf{x}_{\parallel }^{\prime }$. With the mode functions (%
\ref{COS}), the thermal part of the Hadamard function, given by (\ref{WF1}),
is expressed as 
\begin{eqnarray}
G_{T}(x,x^{\prime }) &=&\frac{q(ww^{\prime })^{D/2}}{(2\pi )^{D-3}a^{D-1}L}%
\sum_{\sigma }\frac{p\lambda }{E_{\sigma }}e^{inq\Delta \phi +ik_{l}\Delta
z+i\mathbf{k}\cdot \Delta \mathbf{x}_{\parallel }}J_{q|n+\alpha |}(\lambda r)
\notag \\
&&\times J_{q|n+\alpha |}(\lambda r^{\prime })J_{\nu }(pw)J_{\nu
}(pw^{\prime })\sum_{u=\pm }\frac{e^{-uiE_{\sigma }\Delta t}}{e^{\beta
(E_{\sigma }-u\mu )}-1}.  \label{wight_T}
\end{eqnarray}%
In the discussion below we will be mainly concerned with the finite
temperature effects.

Here it should be noted that in order to have positive-definite values for
the particle and antiparticle numbers the condition $|\mu |\leq E_{0}$ is
required, where $E_{0}$ is the minimal value of the energy. Assuming that $|%
\tilde{\delta}|<1/2$, the minimal energy in the problem at hand corresponds
to the modes with $(p,\lambda ,l,\mathbf{k})=(0,0,0,\mathbf{0})$ and it is
given by%
\begin{equation}
E_{0}=\frac{2\pi }{L}|\delta _{0}|\ .  \label{E0}
\end{equation}%
This corresponds to the minimal Kaluza-Klein mass. With this value for the
minimal energy of the modes, the region for allowed values of the chemical
potential is specified by the condition%
\begin{equation}
|\mu |\leq \frac{2\pi }{L}|\delta _{0}|.  \label{mucond}
\end{equation}%
In decompactified models or in compactified models with $\delta _{0}=0$ one
has $E_{0}=0$ and the chemical potential has to be set equal to zero. Here
we should point out an important difference between the problems in the
Minkowski and AdS bulks. In the Minkowskian problem, for the minimal value
of the energy one has $E_{\mathrm{(M)}0}=\sqrt{E_{0}^{2}+m^{2}}$ with $E_{0}$
from (\ref{E0}). In the models with $L\rightarrow \infty $ or $\delta _{0}=0$
we get $E_{\mathrm{(M)}0}=m$ and the chemical potential for massive fields
needs not be zero: the values in the range $|\mu |\leq m$ are allowed. The
reason for the mentioned difference between the Minkowski and AdS
backgrounds is that the energy in the second case does not depend on the
mass.

At this stage it is worth to mention about boundary conditions on the AdS
boundary. The general solution of the field equation has the form similar to
(\ref{COS}) with the Bessel function $J_{\nu }(pw)$ replaced by a linear
combination of the Bessel and Neumann functions. In the range $\nu \geq 1$
the requirement of the normalizability of the modes excludes the Neumann
function and they are given by (\ref{COS}). In the region $0\leq \nu <1$,
both the functions are allowed and the additional coefficient in the linear
combination should be fixed by imposing a boundary condition on the AdS
boundary. Our choice in (\ref{COS}) corresponds to the Dirichlet condition.
The general class of Robin-type conditions has been discussed in \cite%
{Ishi04}-\cite{Morl21}.

In the context of the AdS/CFT correspondence it is also of interest to
consider the bulk-to-boundary propagator. It plays an important role in the
map between observables of the dual theories. For the pure AdS bulk the
propagator has been discussed in the papers \cite{Witt98b}. The models with
additional boundaries were considered in \cite{Alis11,Beze15Ort}. The
evaluation of the bulk-to-boundary propagator is usually realized in
Euclidean signature. Introducing the set of coordinates $\mathbf{X}%
_{\parallel }=(X^{0},\mathbf{x}_{\parallel })$, the respective line element
in the problem at hand reads%
\begin{equation}
ds_{\mathrm{E}}^{2}=\left( \frac{a}{w}\right) ^{2}\left( dr^{2}+r^{2}d\phi
^{2}+dz^{2}+d\mathbf{X}_{\parallel }^{2}+dw^{2}\right) \ .  \label{ds2E}
\end{equation}%
With this metric tensor, for the solutions of the field equation, which do
not diverge in the limit $w\rightarrow \infty $, the $w$-dependence is
expressed in terms of the Macdonald function $K_{\nu }(\bar{p}w)$ and they
are given by 
\begin{equation}
\varphi _{\mathrm{E}\sigma }(x)=\mathrm{const}\cdot w^{\frac{D}{2}}K_{\nu }(%
\bar{p}w)J_{q|n+\alpha |}(\lambda r)e^{iqn\phi +ik_{l}z+i\mathbf{K}\cdot 
\mathbf{X}_{\parallel }}\ ,  \label{phiE}
\end{equation}%
where $\bar{p}=\sqrt{\lambda ^{2}+\tilde{k}_{l}^{2}+\mathbf{K}^{2}}$. The
general solution is presented in the form of the expansion%
\begin{eqnarray}
\varphi (\mathbf{X}_{D},w) &=&w^{D/2}\int d\mathbf{K}\int_{0}^{\infty
}d\lambda \sum_{n,l=-\infty }^{+\infty }\varphi _{n,l}(\lambda ,\mathbf{K}) 
\notag \\
&&\times \bar{p}^{\nu }K_{\nu }(\bar{p}w)J_{q|n+\alpha |}(\lambda
r)e^{iqn\phi +ik_{l}z+i\mathbf{K}\cdot \mathbf{X}_{\parallel }},
\label{phiExp}
\end{eqnarray}%
with the coordinates in $D$-dimensional space $\mathbf{X}_{D}=(r,\phi ,z,%
\mathbf{X}_{\parallel })$ and expansion coefficients $\varphi _{n,l}(\lambda
,\mathbf{K})$.

Inverting (\ref{phiExp}), it can be seen that 
\begin{equation}
\varphi _{n,l}(\lambda ,\mathbf{K})=\frac{\left( 2\pi \right) ^{3-D}\lambda 
}{2^{\nu -1}\phi _{0}L\Gamma \left( \nu \right) }\int d\mathbf{X}%
_{D}\,\varphi _{(0)}(\mathbf{X}_{D})J_{q|n+\alpha |}(\lambda r)e^{-iqn\phi
-ik_{l}z-i\mathbf{K}\cdot \mathbf{X}_{\parallel }},  \label{phinl}
\end{equation}%
where $\varphi _{(0)}(\mathbf{X}_{D})=\lim_{w\rightarrow 0}w^{\nu
-D/2}\varphi (\mathbf{X}_{D},w)$ and 
\begin{equation}
\int d\mathbf{X}_{D}=\int d\mathbf{X}_{\parallel }\mathbf{\,}%
\int_{0}^{\infty }dr\,r\int_{0}^{\phi _{0}}d\phi \,\int_{0}^{L}dz.
\label{IntXD}
\end{equation}%
Inserting this back into the expansion (\ref{phiExp}) we get%
\begin{equation}
\varphi (\mathbf{X}_{D},w)=\int d\mathbf{X}_{D}^{\prime }G(\mathbf{X}_{D};%
\mathbf{X}_{D}^{\prime },w)\varphi _{(0)}(\mathbf{X}_{D}^{\prime }),
\label{phiD}
\end{equation}%
where the bulk-to-boundary propagator is given by%
\begin{eqnarray}
G(\mathbf{X}_{D};\mathbf{X}_{D}^{\prime },w) &=&\frac{\left( 2\pi \right)
^{3-D}w^{D/2}}{2^{\nu -1}\phi _{0}L\Gamma \left( \nu \right) }\int d\mathbf{K%
}\int_{0}^{\infty }d\lambda \sum_{n,l=-\infty }^{+\infty }\,\bar{p}^{\nu
}K_{\nu }(\bar{p}w)  \notag \\
&&\times \lambda J_{q|n+\alpha |}(\lambda r)J_{q|n+\alpha |}(\lambda
r^{\prime })e^{iqn\Delta \phi +ik_{l}\Delta z+i\mathbf{K}\cdot \Delta 
\mathbf{X}_{\parallel }},  \label{Gprop}
\end{eqnarray}%
with $\Delta \phi =\phi -\phi ^{\prime }$, $\Delta z=z-z^{\prime }$, and $%
\Delta \mathbf{X}_{\parallel }=\mathbf{X}_{\parallel }-\mathbf{X}_{\parallel
}^{\prime }$.

\section{Charge and current densities: General expressions}

\label{Current}

The current density operator for a scalar field is given by 
\begin{equation}
{j}_{s}=ie\left[ {\hat{\varphi}}^{\dagger }(x)D_{s}{\hat{\varphi}}(x)-{\hat{%
\varphi}}\left( D_{s}{\hat{\varphi}}\right) ^{\dagger }\right] \ .
\label{C2}
\end{equation}%
The corresponding thermal average is obtained in terms of the Hadamard
function as 
\begin{equation}
\langle {j}_{s}\rangle =\frac{ie}{2}\ \mathrm{lim}_{x^{\prime }\rightarrow x}%
\left[ (\partial _{s}-\partial _{s}^{\prime }+2ieA_{s})G(x,x^{\prime })%
\right] \ .  \label{Cur}
\end{equation}%
Because the Hadamard function is decomposed as shown in (\ref{W_total}), the
same will happen for the expectation value of the current density: 
\begin{equation}
\left\langle j_{s}\right\rangle =\left\langle j_{s}\right\rangle
^{(0)}+\left\langle j_{s}\right\rangle ^{(T)}\ ,  \label{Curr}
\end{equation}%
where $\left\langle j_{s}\right\rangle ^{(0)}$ corresponds to the zero
temperature current, already investigated in \cite{Oliv19}, and $%
\left\langle j_{s}\right\rangle ^{(T)}$ is the contribution from particles
and antiparticles. Here we are interested in the latter contribution. In the
limit $T\rightarrow 0$ we expect that the thermal part will vanish and only
the first term, the zero temperature contribution, survives.

\subsection{Charge density}

We start our investigation of thermal expectation values from the charge
density, $\left\langle j_{0}\right\rangle ^{(T)}$\footnote{%
In \cite{Oliv19}, it has been shown that the renormalized zero temperature
charge density vanishes.}. Knowing that $A_{0}=0$, and substituting the
thermal Hadamard function in the corresponding expression, we get 
\begin{eqnarray}
\left\langle j^{0}\right\rangle ^{(T)} &=&\frac{eqw^{D+2}}{(2\pi
)^{D-3}a^{D+1}L}\sum_{l,n=-\infty }^{+\infty }\int d\mathbf{k}%
\int_{0}^{\infty }dp\,pJ_{\nu }^{2}(pw)  \notag \\
&&\times \int_{0}^{\infty }d\lambda \lambda \,J_{q|n+\alpha |}^{2}(\lambda
r)\sum_{u=\pm }u\frac{1}{e^{\beta (E_{\sigma }-u\mu )}-1}\ .
\label{Chargeden}
\end{eqnarray}%
As we can observe from the above expression, the thermal charge density is
an odd function of $\mu $. So, when the chemical potential is zero the
contributions from the particles and antiparticles cancel each other and the
total charge density vanishes. The chemical potential has opposite signs for
particles and antiparticles and its nonzero value imbalances the
particle-antiparticle contributions.

Because the summation over $n$ goes from $-\infty $ to $\infty $, the
expectation value (\ref{Chargeden}) is an even periodic function of the
parameter $\alpha $ with the period equal to 1. Writing $\alpha $ in the
form 
\begin{equation}
\alpha =N_{\alpha }+\alpha _{0}\ ,  \label{Aharonov}
\end{equation}%
being $N_{\alpha }$ an integer number and $\alpha _{0}$ the fractional part, 
$|\alpha _{0}|\leq 1/2$, we observe that the charge density depends only on
the fractional part $\alpha _{0}$. Bearing in mind that the parameter $%
\alpha $ is related to the magnetic flux confined inside the core of the
defect, this feature is interpreted as an Aharonov-Bohm type effect. In a
similar way, the charge density is an even periodic function of $\tilde{%
\delta}$, again, with the period 1. This corresponds to the periodicity with
respect to the magnetic flux enclosed by compact dimension with the period
equal to flux quantum. Presenting%
\begin{equation}
\tilde{\delta}=N_{\delta }+\delta _{0}\,,\;|\delta _{0}|\leq 1/2,
\label{delcomp}
\end{equation}%
we see that the charge density does not depend on integer $N_{\delta }$. As
it has been already discussed, for $\delta _{0}=0$ one should take $\tilde{%
\mu}=0$ and in this case both the zero temperature and thermal charge
densities vanish.

Assuming the condition $|\mu |\leq E_{0}$, we can employ the series
expansion 
\begin{equation}
(e^{y}-1)^{-1}=\sum_{j=1}^{\infty }e^{-jy}\ ,  \label{Expansion}
\end{equation}%
in (\ref{Chargeden}). This gives 
\begin{eqnarray}
\left\langle j^{0}\right\rangle ^{(T)} &=&\frac{2eqw^{D+2}}{(2\pi
)^{D-3}a^{D+1}L}\sum_{n=-\infty }^{\infty }\sum_{l=-\infty }^{\infty }\int d%
\mathbf{k}\int_{0}^{\infty }dpp\int_{0}^{\infty }d\lambda \lambda  \notag \\
&\times &J_{q|n+\alpha |}^{2}(\lambda r)J_{\nu }^{2}(pw)\sum_{j=1}^{\infty
}e^{-j\beta E_{\sigma }}\sinh (j\beta {\mu })\ .  \label{Chargeden1}
\end{eqnarray}%
For the further transformation of this expression we use the relation 
\begin{equation}
e^{-j\beta E_{\sigma }}=\frac{j\beta }{\sqrt{\pi }}\int_{0}^{\infty }\frac{du%
}{u^{2}}e^{-E_{\sigma }^{2}u^{2}-j^{2}\beta ^{2}/(4u^{2})}\ ,  \label{IntRep}
\end{equation}%
with $E_{\sigma }$ from (\ref{Energy}). Substituting this in (\ref%
{Chargeden1}), the integral over the momentum $\mathbf{k}$ gives $\int d%
\mathbf{k}\,e^{-\mathbf{k}^{2}u^{2}}=\pi ^{D/2-2}u^{4-D}$, whereas the
integrals over $\lambda $ and $p$ are evaluated with the help of the formula 
\cite{Grad}: 
\begin{equation}
\int_{0}^{\infty }dx\ xe^{-\alpha ^{2}x^{2}}J_{\nu }(\sigma x)J_{\nu
}(\sigma ^{\prime }x)=\frac{1}{2\alpha ^{2}}\exp \left( -\frac{\sigma
^{2}+\sigma ^{\prime 2}}{4\alpha ^{2}}\right) I_{\nu }\left( \frac{\sigma
\sigma ^{\prime }}{2\alpha ^{2}}\right) \ ,  \label{IntForm}
\end{equation}%
being $I_{\nu }(z)$ the modified Bessel function \cite{Abra}. Passing to a
new integration variable $x=w^{2}/2u^{2}$, the charge density is transformed
to%
\begin{eqnarray}
\left\langle j^{0}\right\rangle ^{(T)} &=&\frac{2^{\frac{3-D}{2}}e\beta w}{%
\pi ^{(D-1)/2}a^{D+1}L}\int_{0}^{\infty }dx\,x^{\frac{D-1}{2}}e^{-x}I_{\nu
}(x){\mathcal{J}}_{0}(q,\alpha _{0},x\rho ^{2})  \notag \\
&&\times \sum_{l=-\infty }^{\infty }e^{-\tilde{k}_{l}^{2}w^{2}/(2x)}%
\sum_{j=1}^{\infty }j\sinh (j\beta {\mu })e^{-j^{2}\beta ^{2}x/(2w^{2})}\ ,
\label{j02}
\end{eqnarray}%
where we have introduced the notations%
\begin{equation}
\rho =r/w,  \label{ro}
\end{equation}%
and%
\begin{equation}
{\mathcal{J}}_{0}(q,\alpha _{0},z)=qe^{-z}\sum_{n=-\infty }^{+\infty }\
I_{q|n+\alpha |}(z).  \label{Ical}
\end{equation}%
Note that ${\mathcal{J}}_{0}(1,0,z)=1$.

An alternative representation for the series over $l$ is obtained by using
the Poison summation formula%
\begin{equation}
\sum_{l=-\infty }^{\infty }g(l\alpha )=\frac{1}{\alpha }\sum_{l=-\infty
}^{\infty }\tilde{g}(2\pi l/\alpha ),  \label{Pois}
\end{equation}%
for a given function $g(x)$ with the Fourier transform $\tilde{g}%
(y)=\int_{-\infty }^{+\infty }dxe^{-iyx}g(x)$. Introducing the function%
\begin{equation}
F_{1}(\delta _{0},u)=\sideset{}{'}{\sum}_{l=0}^{\infty }\cos \left( 2\pi
l\delta _{0}\right) e^{-l^{2}u}  \label{F1}
\end{equation}%
we can see that%
\begin{equation}
F_{1}(\delta _{0},u)=\sqrt{\frac{\pi }{4u}}\sum_{l=-\infty }^{\infty }\exp %
\left[ -\frac{\pi ^{2}}{u}(l+\delta _{0})^{2}\right] .  \label{F1b}
\end{equation}%
The prime in (\ref{F1}) means that the term $l=0$ should be taken with an
additional coefficient 1/2. With the notations above, the expression for the
charge density is written in the form%
\begin{equation}
\left\langle j^{0}\right\rangle ^{(T)}=\frac{4ea^{-D-1}}{\left( 2\pi \right)
^{\frac{D}{2}}}\int_{0}^{\infty }dx\,x^{\frac{D}{2}}\frac{I_{\nu }(x)}{e^{x}}%
{\mathcal{J}}_{0}(q,\alpha _{0},x\rho ^{2})F_{1}\left( \delta _{0},\frac{%
L^{2}x}{2w^{2}}\right) \partial _{{\mu }}F_{2}\left( \beta {\mu },\frac{%
\beta ^{2}x}{2w^{2}}\right) \ ,  \label{j03}
\end{equation}%
where we have introduced a new function%
\begin{equation}
F_{2}(\gamma ,u)=\sum_{j=1}^{\infty }\cosh \left( j\gamma \right)
e^{-j^{2}u}.  \label{F2}
\end{equation}%
An equivalent representation for the function (\ref{F2}) is obtained on the
base of the resummation formula (\ref{Pois}):%
\begin{equation}
F_{2}(\gamma ,u)=\sqrt{\frac{\pi }{u}}\exp \left( \frac{\gamma ^{2}}{4u}%
\right) \sideset{}{'}{\sum}_{j=0}^{\infty }e^{-\pi ^{2}j^{2}/u}\cos \left(
\pi j\frac{{\gamma }}{u}\right) -\frac{1}{2}.  \label{F2b}
\end{equation}%
We have the relation%
\begin{equation}
F_{1}(\delta _{0},u)=\frac{1}{2}+F_{2}(2\pi i\delta _{0},u)  \label{RelF12}
\end{equation}%
Note that the functions (\ref{F1}) and (\ref{F2}) are expressed in terms of
the Jacobi theta function $\vartheta _{3}(z,u)$ \cite{Abra}:%
\begin{eqnarray}
F_{1}(\delta _{0},u) &=&\frac{1}{2}\vartheta _{3}(\pi \delta _{0},e^{-u}), 
\notag \\
F_{2}(\gamma ,u) &=&\sqrt{\frac{\pi }{4u}}\exp \left( \frac{\gamma ^{2}}{4u}%
\right) \vartheta _{3}\left( \frac{\pi \gamma }{2u},e^{-\frac{\pi ^{2}}{u}%
}\right) -\frac{1}{2}  \label{F12Jac}
\end{eqnarray}

For the function (\ref{Ical}) we can use the integral representation \cite%
{Brag15}: 
\begin{equation}
{\mathcal{J}}_{0}(q,\alpha _{0},z)=2\sideset{}{'}{\sum}_{k=0}^{[q/2]}\cos
\left( 2\pi k\alpha _{0}\right) e^{-2zs_{k}^{2}}-\frac{q}{\pi }%
\int_{0}^{\infty }dy\ \frac{e^{-2z\cosh ^{2}({y/2)}}h_{0}(q,\alpha _{0},y)}{%
\cosh (qy)-\cos (q\pi )}\ ,  \label{Sum_I}
\end{equation}%
with the notations 
\begin{equation}
s_{k}=\sin (\pi k/q),  \label{sk}
\end{equation}%
and%
\begin{equation}
h_{0}(q,\alpha _{0},y)=\sin \left[ \left( 1-|\alpha _{0}|\right) q\pi \right]
\cosh \left( |\alpha _{0}|qy\right) +\sin \left( |\alpha _{0}|q\pi \right)
\cosh \left[ \left( 1-|\alpha _{0}|\right) qy\right] .  \label{h}
\end{equation}%
In (\ref{Sum_I}), $[q/2]$ represents the integer part of $q/2$ and the prime
on the summation means that the term $k=0$ and the term $k=q/2$ for even
values of $q$ should be taken with the coefficient $1/2$. Combining (\ref%
{j03}) and (\ref{Sum_I}), we obtain an alternative representation of the
thermal charge density.

The contribution coming from the $k=0$ term in (\ref{Sum_I}) corresponds to
the charge density in the geometry where the cosmic string is absent ($q=1$, 
$\alpha _{0}$). In this case the charge distribution is homogeneous with
respect to the radial coordinate and we will denote the corresponding
density by $\left\langle j^{0}\right\rangle _{0}^{(T)}$. The expression for
that part is obtained from (\ref{j03}) substituting ${\mathcal{J}}%
_{0}(q,\alpha _{0},x\rho ^{2})\rightarrow {\mathcal{J}}_{0}(1,0,x\rho
^{2})=1 $. The radial inhomogeneity in the charge distribution is a
consequence of the presence of the cosmic string. The total charge $\delta
Q_{\mathrm{cs}}$ in the volume element $d\mathbf{x}_{\parallel }dw$ of the
subspace $(\mathbf{x}_{\parallel },w)$, induced by the cosmic string is
finite and is obtained by integrating the difference $\left\langle
j^{0}\right\rangle ^{(T)}-\left\langle j^{0}\right\rangle _{0}^{(T)}$ with
respect to the coordinates $r$, $\phi $ and $z$. The integrals over $\phi $
and $z$ give a factor $2\pi L/q$ and we get%
\begin{equation}
\delta Q_{\mathrm{cs}}=\frac{2\pi L}{q}\left( \frac{a}{w}\right) ^{D+1}d%
\mathbf{x}_{\parallel }dw\int_{0}^{\infty }dr\,r\left[ \left\langle
j^{0}\right\rangle ^{(T)}-\left\langle j^{0}\right\rangle _{0}^{(T)}\right] .
\label{Qcs}
\end{equation}%
For evaluation of the radial integral it is convenient to use the
representation (\ref{Ical}) for the function ${\mathcal{J}}_{0}(q,\alpha
_{0},z)$ in (\ref{j03}). The integral is reduced to 
\begin{equation}
\lim_{p\rightarrow 1+}\int_{0}^{\infty }dr\,re^{-px\rho ^{2}}\left[
qI_{q|n+\alpha _{0}|}(x\rho ^{2})-I_{|n|}(x\rho ^{2})\right] .  \label{Intr}
\end{equation}%
Here, we have introduced the factor $p>1$ in the exponent in order to have a
possibility to integrate the separate parts by using the formula from \cite%
{Prud2} (for $p=1$ these separate integrals diverge). After evaluating the
integral in (\ref{Intr}), the series over $n$ is reduced to the sum of
geometric progression. Taking the limit $p\rightarrow 1+$ at the end, one
finds%
\begin{equation}
\sum_{n=-\infty }^{+\infty }\int_{0}^{\infty }dr\,re^{-x\rho ^{2}}\left[
qI_{q|n+\alpha _{0}|}(x\rho ^{2})-I_{|n|}(x\rho ^{2})\right] =\frac{w^{2}}{2x%
}\left[ \frac{q^{2}-1}{6}-q^{2}|\alpha _{0}|\left( 1-|\alpha _{0}|\right) %
\right] .  \label{Limsum}
\end{equation}%
Hence, for the total charge induced by the cosmic string, per unit volume in
the subspace $(\mathbf{x}_{\parallel },w)$, we obtain%
\begin{eqnarray}
\frac{\delta Q_{\mathrm{cs}}}{d\mathbf{x}_{\parallel }dw} &=&\frac{2eLw^{1-D}%
}{\left( 2\pi \right) ^{\frac{D}{2}-1}}\left[ \frac{q^{2}-1}{6q}-q|\alpha
_{0}|\left( 1-|\alpha _{0}|\right) \right]  \notag \\
&&\times \int_{0}^{\infty }dx\,x^{\frac{D}{2}-1}\frac{I_{\nu }(x)}{e^{x}}%
F_{1}\left( \delta _{0},\frac{L^{2}x}{2w^{2}}\right) \partial _{{\mu }%
}F_{2}\left( \beta {\mu },\frac{\beta ^{2}x}{2w^{2}}\right) \ ,  \label{Qcsd}
\end{eqnarray}%
Note that the part containing the parameters of the cosmic string is
factorized. The corresponding factor can be either positive or negative.
Hence, depending on the planar angle deficit and on the magnetic flux
confined in the cosmic string core, its presence can either increase or
decrease the total charge.

\subsection{Azimuthal current density}

Now we turn to the spatial components of the current density. First of all
we can see that the components in (\ref{Cur}) with $s=1,4,\ldots ,D$ become
zero. By using the Hadamard function (\ref{wight_T}), for the finite
temperature contribution to the expectation value of the covariant component
of the current density along the azimuthal direction, $\langle j^{2}\rangle
^{(T)}=-\langle j_{2}\rangle ^{(T)}/(a\rho )^{2}$, we obtain 
\begin{eqnarray}
\langle j^{2}\rangle ^{(T)} &=&\frac{eq^{2}w^{D+2}a^{-1-D}}{2(2\pi
)^{D-3}Lr^{2}}\sum_{l,n=-\infty }^{+\infty }(n+\alpha )\int d\mathbf{k}%
\int_{0}^{\infty }dp\,pJ_{\nu }^{2}(pw)  \notag \\
&&\times \int_{0}^{\infty }d\lambda \,\frac{\lambda }{E_{\sigma }}%
J_{q|n+\alpha |}^{2}(\lambda r)\ \sum_{u=\pm }\frac{1}{e^{\beta (E_{\sigma
}+u\mu )}-1}\ .  \label{J_Azim}
\end{eqnarray}%
Note that this component is an odd periodic function of the parameter $%
\alpha $, consequently in the absence of the magnetic flux along the string
it vanishes. Moreover, it is an even function of the chemical potential $\mu 
$ and an even periodic function of the flux $\Phi _{3}$. For zero chemical
potential the contributions from the particles and antiparticles coincide.

Using the expansion (\ref{Expansion}) we can write (\ref{J_Azim}) in the
form 
\begin{eqnarray}
\langle j^{2}\rangle ^{(T)} &=&\frac{eq^{2}w^{D+2}}{(2\pi
)^{D-3}a^{D+1}Lr^{2}}\sum_{n=-\infty }^{\infty }(n+\alpha )\sum_{l=-\infty
}^{\infty }\int_{0}^{\infty }d\lambda \,\lambda J_{q|n+\alpha |}^{2}(\lambda
r)  \notag \\
&&\times \int_{0}^{\infty }dp\,pJ_{\nu }^{2}(pw)\int d\mathbf{k}%
\sum_{j=1}^{\infty }\frac{e^{-j\beta E_{\sigma }}}{E_{\sigma }}\cosh (j\beta
\mu )\ .  \label{J_Azim1}
\end{eqnarray}%
The further steps are similar to those we have employed for the charge
density. With the help of the relation 
\begin{equation}
\frac{e^{-j\beta E_{\sigma }}}{E_{\sigma }}=\frac{2}{\sqrt{\pi }}%
\int_{0}^{\infty }du\,e^{-E_{\sigma }^{2}u^{2}-j^{2}\beta ^{2}/(4u^{2})},
\label{Rel2}
\end{equation}%
the integrals over $\lambda $ and $p$ are evaluated by using the formula (%
\ref{IntForm}). The current density is presented as%
\begin{equation}
\langle j^{2}\rangle ^{(T)}=\frac{2ea^{-1-D}}{\left( 2\pi \right) ^{\frac{D}{%
2}}}\int_{0}^{\infty }dx\,x^{\frac{D}{2}}\frac{I_{\nu }\left( x\right) }{%
e^{x}}{\mathcal{J}}_{2}(q,\alpha _{0},x\rho ^{2})F_{1}\left( \delta _{0},%
\frac{L^{2}x}{2w^{2}}\right) F_{2}\left( \beta {\mu },\frac{\beta ^{2}x}{%
2w^{2}}\right) \ ,  \label{j2}
\end{equation}%
with the notation%
\begin{equation}
{\mathcal{J}}_{2}(q,\alpha _{0},z)=q^{2}\frac{e^{-z}}{z}\sum_{n=-\infty
}^{+\infty }\ (n+\alpha )I_{q|n+\alpha |}(z),  \label{Fcal}
\end{equation}%
and with the functions defined by (\ref{F1b}) and (\ref{F2}).

An alternative expression for the azimuthal current density is obtained by
using the representation \cite{Brag15}%
\begin{eqnarray}
{\mathcal{J}}_{2}(q,\alpha _{0},z) &=&2\sideset {}{'}\sum_{k=1}^{[q/2]}\sin {%
(2\pi k/q)}\sin \left( 2\pi k\alpha _{0}\right) e^{-2zs_{k}^{2}}  \notag \\
&&-\frac{q}{\pi }\int_{0}^{\infty }dy\ \frac{e^{-2z\cosh ^{2}{(y/2)}%
}h_{2}(q,\alpha _{0},y)}{\cosh (qy)-\cos (q\pi )}\ ,  \label{Kcal}
\end{eqnarray}%
where 
\begin{equation}
h_{2}(q,\alpha _{0},y)=\sinh y\left\{ \sinh \left( q\alpha _{0}y\right) \sin 
\left[ \left( 1-|\alpha _{0}|\right) q\pi \right] -\sin \left( q\alpha
_{0}\pi \right) \sinh \left[ \left( 1-|\alpha _{0}|\right) qy\right]
\right\} \ .  \label{g}
\end{equation}%
Of course, ${\mathcal{J}}_{2}(q,0,z)=0$ and, as it has been mentioned above,
for $\alpha _{0}=0$ the azimuthal current density vanishes. In addition, one
has ${\mathcal{J}}_{2}(q,\pm 1/2,z)=0$ and the azimuthal current density is
zero for $\alpha _{0}=\pm 1/2$ as well.

\subsection{Axial current}

It remains to consider the component of the current density along the
compact dimension $z$, corresponding to $s=3$ in (\ref{Cur}). Substituting
the Hadamard function (\ref{W_total}) and using the definition (\ref{Phase}%
), for the contravariant component of the thermal part we get 
\begin{eqnarray}
\langle j^{3}\rangle ^{(T)} &=&\frac{eqw^{D+2}}{(2\pi )^{D-3}a^{D+1}L}%
\sum_{l,n=-\infty }^{\infty }\tilde{k}_{l}\int d\mathbf{k}\int_{0}^{\infty
}dp\,pJ_{\nu }^{2}(pw)  \notag \\
&&\times \int_{0}^{\infty }d\lambda \frac{\lambda }{E_{\sigma }}%
J_{q|n+\alpha |}^{2}(\lambda r)\sum_{u=\pm }\frac{1}{e^{\beta (E_{\sigma
}+u\mu )}-1}\ .  \label{j_Ax1}
\end{eqnarray}%
With the help of the expansion (\ref{Expansion}), this gives 
\begin{eqnarray}
\langle j^{3}\rangle ^{(T)} &=&\frac{2eqw^{D+2}}{(2\pi )^{D-3}a^{D+1}L}%
\sum_{n=-\infty }^{\infty }\sum_{l=-\infty }^{\infty }{\tilde{k}}%
_{l}\int_{0}^{\infty }d\lambda \,\lambda J_{q|n+\alpha |}^{2}(\lambda r) 
\notag \\
&&\times \int_{0}^{\infty }dp\,pJ_{\nu }^{2}(pw)\int d\mathbf{k}\
\sum_{j=1}^{\infty }\frac{e^{-j\beta E_{\sigma }}}{E_{\sigma }}\cosh (j\beta
\mu )\ .  \label{j_Ax2}
\end{eqnarray}%
By using the integral representation (\ref{Rel2}) the $\lambda $- and $p$%
-integrals are evaluated with the help of formula (\ref{IntForm}).
Introducing a new integration variable $x=w^{2}/(2u^{2})$, the series over $%
l $ is presented in the form $\sum_{l=-\infty }^{\infty }{\tilde{k}}_{l}e^{-%
\tilde{k}_{l}^{2}w^{2}/(2x)}$. By making use of the resummation formula (\ref%
{Pois}) this series is expressed in terms of the function (\ref{F1}): 
\begin{equation}
\sum_{l=-\infty }^{\infty }{\tilde{k}}_{l}e^{-\tilde{k}_{l}^{2}w^{2}/(2x)}=-%
\frac{L^{2}x^{3/2}}{\sqrt{2}\pi ^{3/2}w^{3}}\partial _{\delta
_{0}}F_{1}\left( \delta _{0},\frac{L^{2}x}{2w^{2}}\right) .  \label{Serl}
\end{equation}%
As a result, the following representation is obtained: 
\begin{equation}
\langle j^{3}\rangle ^{(T)}=-\frac{4eLa^{-D-1}}{\left( 2\pi \right) ^{\frac{D%
}{2}+1}}\int_{0}^{\infty }dx\,x^{\frac{D}{2}}\frac{I_{\nu }\left( x\right) }{%
e^{x}}{\mathcal{J}}_{0}(q,\alpha _{0},x\rho ^{2})F_{2}\left( \beta {\mu },%
\frac{\beta ^{2}x}{2w^{2}}\right) \partial _{\delta _{0}}F_{1}\left( \delta
_{0},\frac{L^{2}x}{2w^{2}}\right) ,  \label{j3}
\end{equation}%
where the function ${\mathcal{J}}_{0}(q,\alpha _{0},z)$ is defined by (\ref%
{Ical}). An equivalent expression for the axial current density is obtained
by using the representation (\ref{Sum_I}). The axial current density is an
even periodic function of the magnetic flux $\Phi _{2}$ with the period of
flux quantum and an odd periodic function of the flux $\Phi _{3}$ with the
same period. In particular, it vanishes for the case $\delta _{0}=0$.

\subsection{Combined formulas}

Introducing a new integration variable $u=x/(2w^{2})$, the expressions for
the thermal contributions to the expectation values of the charge and
current densities are combined in the single formula%
\begin{eqnarray}
\left\langle j^{s}\right\rangle ^{(T)} &=&\frac{2ea^{-D-1}}{\left( 2\pi
\right) ^{\frac{D}{2}}}\int_{0}^{\infty }dx\,x^{\frac{D}{2}}\frac{I_{\nu }(x)%
}{e^{x}}{\mathcal{J}}_{s}(q,\alpha _{0},x\rho ^{2})  \notag \\
&&\times \left( \frac{-L}{\pi }\partial _{\delta _{0}}\right) ^{\delta
_{s3}}F_{1}\left( \delta _{0},\frac{L^{2}x}{2w^{2}}\right) \left( 2\partial
_{{\mu }}\right) ^{\delta _{s0}}F_{2}\left( \beta {\mu },\frac{\beta ^{2}x}{%
2w^{2}}\right) \ ,  \label{jmu}
\end{eqnarray}%
where $s=0,2,3$, ${\mathcal{J}}_{3}(q,\alpha _{0},x)={\mathcal{J}}%
_{0}(q,\alpha _{0},x)$, and the functions ${\mathcal{J}}_{0}(q,\alpha
_{0},x) $, ${\mathcal{J}}_{2}(q,\alpha _{0},x)$ are defined by (\ref{Ical})
and (\ref{Fcal}). Alternative representation is obtained from (\ref{jmu}) by
using (\ref{Sum_I}) and (\ref{Kcal}).

To see the convergence properties of the $x$-integral in (\ref{jmu}) we can
use the asymptotic expressions%
\begin{eqnarray}
F_{1}\left( \delta _{0},u\right) &\approx &\frac{1}{2}+e^{-u}\cos \left(
2\pi \delta _{0}\right) ,  \notag \\
F_{2}\left( \gamma ,u\right) &\approx &e^{-u}\cosh \gamma ,  \label{S2as}
\end{eqnarray}%
for $u\gg 1$ and%
\begin{eqnarray}
F_{1}\left( \delta _{0},u\right) &\approx &\sqrt{\frac{\pi }{4u}}\exp \left(
-\pi ^{2}\frac{\delta _{0}^{2}}{u}\right) ,  \notag \\
F_{2}\left( \gamma ,u\right) &\approx &\sqrt{\frac{\pi }{4u}}\exp \left( 
\frac{{\gamma }^{2}}{4u}\right) ,  \label{S2as2}
\end{eqnarray}%
for $u\ll 1$. For the function ${\mathcal{J}}_{0}(q,\alpha _{0},z)$ one has $%
\lim_{z\rightarrow \infty }{\mathcal{J}}_{0}(q,\alpha _{0},z)=1$. In the
same limit, $z\rightarrow \infty $, the function ${\mathcal{J}}_{2}(q,\alpha
_{0},z)$ behaves like $e^{-2zs_{1}^{2}}$ for $q\geq 2$ and as $e^{-2z}$ for $%
1\leq q<2$. In the opposite limit of small $z$ the corresponding asymptotics
are found from the representations (\ref{Ical}) and (\ref{Fcal}): 
\begin{equation}
{\mathcal{J}}_{s}(q,\alpha _{0},z)\approx \left( \frac{q\alpha _{0}}{z}%
\right) ^{\delta _{s2}}\frac{\ q(z/2)^{q|\alpha _{0}|}}{\Gamma (q|\alpha
_{0}|+1)},\;z\ll 1.  \label{Jcalsm}
\end{equation}%
By using these asymptotics we can see that for the components $s=0,3$ the
integrand in (\ref{jmu}) behaves as 
\begin{equation}
x^{\frac{D-1}{2}}\exp \left( -\frac{\beta ^{2}+L^{2}\delta _{s3}}{2w^{2}}%
x\right) ,  \label{largeu}
\end{equation}
for large $x$ and like 
\begin{equation}
x^{D/2+\nu +q|\alpha _{0}|-2}\exp \left( \frac{{\mu }^{2}-E_{0}^{2}}{2x}%
w^{2}\right) ,  \label{smallu}
\end{equation}%
for small $x$. For the component $s=2$ an additional exponential factor for
large $x$ comes from the function ${\mathcal{J}}_{2}(q,\alpha _{0},x\rho
^{2})$.

In the problem under consideration, the zero temperature current density $%
\left\langle j^{s}\right\rangle ^{(0)}$ has been investigated in \cite%
{Oliv19}. The corresponding charge density vanishes and the components $%
s=2,3 $, adapted to our notations, can be presented in the combined form%
\begin{equation}
\left\langle j^{s}\right\rangle ^{(0)}=\frac{ea^{-D-1}}{\left( 2\pi \right)
^{\frac{D}{2}}}\int_{0}^{\infty }dx\,x^{\frac{D}{2}}\frac{I_{\nu }\left(
x\right) }{e^{x}}{\mathcal{J}}_{s}(q,\alpha _{0},x\rho ^{2})\left( \frac{-L}{%
\pi }\partial _{\delta _{0}}\right) ^{\delta _{s3}}F_{1}\left( \delta _{0},%
\frac{L^{2}x}{2w^{2}}\right) .  \label{js0}
\end{equation}%
The part (\ref{js0}) coming from the $l=0$ term in the definition (\ref{F1})
of the function $F_{1}\left( \delta _{0},u\right) $ corresponds to the
vacuum current density in the geometry with cosmic string where the $z$%
-direction is not compactified. In that geometry the only nonzero component
corresponds to the azimuthal current ($s=2$). By taking into account the
expression (\ref{F2}) for the function $F_{2}\left( \gamma ,u\right) $, we
see that the formula for the total current density $\left\langle
j^{s}\right\rangle $ is obtained from (\ref{jmu}) by the replacement 
\begin{equation}
F_{2}\left( \beta {\mu },\frac{\beta ^{2}x}{2w^{2}}\right) \rightarrow %
\sideset{}{'}{\sum}_{j=0}^{\infty }\cosh \left( j\beta {\mu }\right) \exp
\left( -j^{2}\frac{\beta ^{2}x}{2w^{2}}\right) ,  \label{F2Repl}
\end{equation}%
where the prime means that the term $j=0$ enters with an additional
coefficient 1/2. The part with that term corresponds to the current density (%
\ref{js0}).

The thermal expectation values $\left\langle j^{s}\right\rangle ^{(T)}$ are
the contravariant components in the coordinate system $(t,r,\phi ,z,\mathbf{x%
}_{\parallel })$. On the graphs below we will present the physical
components $\left\langle j^{s}\right\rangle _{\mathrm{p}}^{(T)}$ defined by
the relation $\left\langle j^{s}\right\rangle _{\mathrm{p}}^{(T)}=\sqrt{%
|g_{ss}|}\left\langle j^{s}\right\rangle ^{(T)}$. The dimensionless
quantities $a^{D}\left\langle j^{s}\right\rangle _{\mathrm{p}}^{(T)}$ depend
on the coordinate $w$ through the combinations $r/w$, $L/w$, $\beta /w$, $%
\mu w$. This is a consequence of the maximal symmetry of AdS spacetime. Note
that the proper distance from the cosmic string and the proper length of the
compact dimension, measured by an observer with fixed coordinate $w$, are
given by $ar/w$ and $aL/w$. Hence, the ratios $r/w$ and $L/w$ are the proper
distance from the string and the proper length of the compact dimension
measured in units of the curvature radius $a$. As it has been emphasized
above, the charge density is an odd function of the chemical potential,
whereas the azimuthal and axial current densities are even functions.

In order to see the behavior of the expectation value (\ref{jmu}) near the
AdS boundary and horizon, for fixed values of the other parameters, it is
convenient to introduce a new integration variable $u=x/(2w^{2})$. With this
variable, the dependence of the integrand on $w$ appears through the
function $e^{-2w^{2}x}I_{\nu }(2w^{2}x)$. By using the corresponding
asymptotics we can see that all the components of $\left\langle
j^{s}\right\rangle ^{(T)}$ tend to zero on the AdS boundary like $w^{D+2\nu
+2}$. In the near horizon limit, corresponding to large values of $w$, one
gets $\left\langle j^{s}\right\rangle ^{(T)}\propto w^{D+1}$.

\begin{figure}[tbph]
\begin{center}
\begin{tabular}{cc}
\epsfig{figure=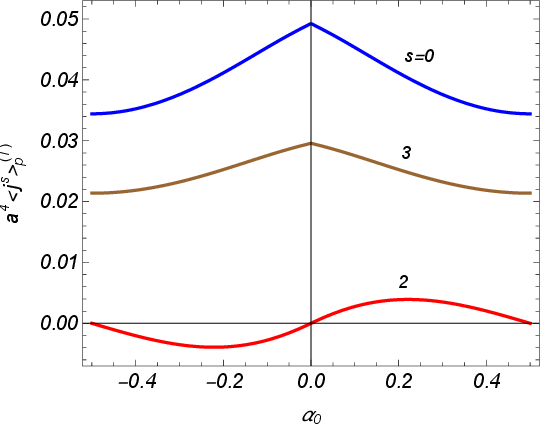,width=7.5cm,height=6cm} & \quad %
\epsfig{figure=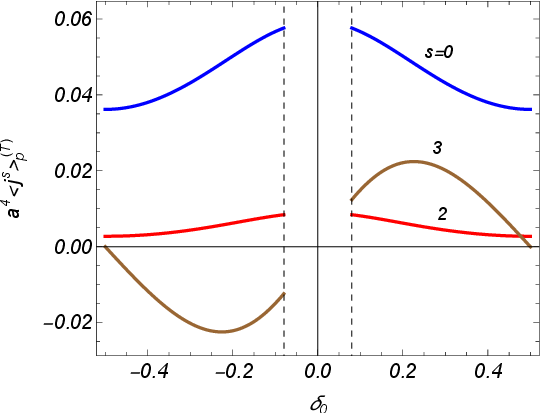,width=7.5cm,height=6cm}%
\end{tabular}%
\end{center}
\caption{The thermal charge and current densities in the model with $D=4$ as
functions of fractional parts of the ratio of magnetic fluxes to the flux
quantum. The graphs are plotted for $r/w=0.5$, $L/w=1$, $wT=1$, $w\protect%
\mu =0.5$, $q=2.5$. For the left panel $\protect\delta _{0}=1/3$ and for the
right one $\protect\alpha _{0}=0.2$.}
\label{fig1}
\end{figure}

The current density is a periodic functions of the magnetic fluxes $\Phi
_{2} $ and $\Phi _{3}$ with the period of flux quantum. The charge density
is an even function of both these fluxes. The azimuthal current is an odd
function of $\Phi _{2}$ and an even function of $\Phi _{3}$. The component $%
\left\langle j^{3}\right\rangle ^{(T)}$ is an even function of $\Phi _{2}$
and an odd function of $\Phi _{3}$. In Figure \ref{fig1} the physical
components of the thermal expectation values of the charge and current
densities are plotted for a minimally coupled massless scalar field ($\nu
=D/2$) as functions of the fractional parts of the magnetic fluxes
determined by the parameters $\alpha _{0}$ and $\delta _{0}$. The graphs are
plotted for the $D=4$ model with fixed values $r/w=0.5$, $L/w=1$, $wT=1$, $%
w\mu =0.5$, $q=2.5$. For the left panel we have taken $\delta _{0}=1/3$ and
for the right panel $\alpha _{0}=0.2$. For given values of the length of the
compact dimension and chemical potential, the allowed values of the
parameter $\delta _{0}$ are restricted by the condition $|\delta _{0}|\leq
|\mu |L/(2\pi )$. The vertical dashed lines on the right panel separate the
regions of those allowed values. Note that all the components are continuous
functions of the parameter $\alpha _{0}$. However, the derivatives of the
charge density and axial current density are discontinuous at $\alpha _{0}=0$%
.

\section{Special cases}

\label{sec:Special}

\subsection{Zero angular deficit}

Let us consider some special cases of general formula (\ref{jmu}). In the
absence of the cosmic string and magnetic flux one has $q=1$ and $\alpha
_{0}=0$ and we will denote the corresponding current density by $%
\left\langle j^{s}\right\rangle _{0}^{(T)}$. By taking into account that ${%
\mathcal{J}}_{2}(1,0,z)=0$ we see that the azimuthal current density
vanishes, $\left\langle j^{2}\right\rangle _{0}^{(T)}=0$. For the charge and
axial current densities, by using ${\mathcal{J}}_{0}(1,0,z)=1$, we get%
\begin{eqnarray}
\left\langle j^{s}\right\rangle _{0}^{(T)} &=&\frac{4e}{\pi ^{\frac{D}{2}}}%
w\left( \frac{w}{a}\right) ^{D+1}\int_{0}^{\infty }du\,u^{\frac{D}{2}}\frac{%
I_{\nu }(2w^{2}u)}{e^{2w^{2}u}}  \notag \\
&&\times \left( \frac{-L}{\pi }\partial _{\delta _{0}}\right) ^{\delta
_{s3}}F_{1}\left( \delta _{0},L^{2}u\right) \left( 2\partial _{{\mu }%
}\right) ^{\delta _{s0}}F_{2}\left( \beta {\mu },\beta ^{2}u\right) \ ,
\label{jmu10}
\end{eqnarray}%
for $s=0,3$. In the absence of planar angle deficit, $q=1$, the expressions
for the functions ${\mathcal{J}}_{s}(q,\alpha _{0},z)$ are simplified to%
\begin{eqnarray}
{\mathcal{J}}_{0}(1,\alpha _{0},z) &=&1-\frac{2}{\pi }\sin \left( \pi
|\alpha _{0}|\right) \int_{0}^{\infty }dy\ \frac{e^{-2z\cosh ^{2}y}}{\cosh y}%
\cosh \left[ \left( 1-2|\alpha _{0}|\right) y\right] \ ,  \notag \\
{\mathcal{J}}_{2}(1,\alpha _{0},z) &=&\frac{4}{\pi }\sin \left( \pi \alpha
_{0}\right) \int_{0}^{\infty }dy\ e^{-2z\cosh ^{2}{y}}\sinh \left[ \left(
1-2|\alpha _{0}|\right) y\right] \sinh y.  \label{Jcal02}
\end{eqnarray}%
The parts coming from the integral terms in (\ref{Jcal02}) correspond to the
contribution of the magnetic flux $\Phi _{2}$.

\subsection{Zero chemical potential}

For the zero chemical potential, ${\mu }{=0}$, the thermal charge density
vanishes $\left\langle j^{0}\right\rangle ^{(T)}=0$ as a consequence of the
cancellation between the contributions coming from the particles and
antiparticles. The expressions for the thermal azimuthal and axial current
densities are obtained from (\ref{jmu}) taking $F_{2}\left( 0,u\right)
=\sum_{j=1}^{\infty }e^{-j^{2}u}$. Another expression for $F_{2}\left(
0,u\right) $ follows from (\ref{F2b}). An equivalent representation is
obtained by substituting in (\ref{jmu}) the expression (\ref{F1}). The $u$%
-integral is evaluated by using the formula \cite{Grad}%
\begin{equation}
\int_{0}^{\infty }dx\,x^{D/2}e^{-bx}I_{\nu }(x)=\sqrt{\frac{2}{\pi }}Z_{\nu
}^{D}(b),\;b>1,  \label{IntZ}
\end{equation}%
with the function in the right-hand side defined by%
\begin{equation}
Z_{\nu }^{D}(u)=e^{-i\pi /2(D+1)}\frac{Q_{\nu -1/2}^{\frac{D+1}{2}}(u)}{%
(u^{2}-1)^{\frac{D+1}{4}}}\ ,  \label{ZD}
\end{equation}%
where $Q_{\gamma }^{\mu }(u)$ represents the associated Legendre function of
the second kind \cite{Abra}. The current densities are presented as%
\begin{eqnarray}
\left\langle j^{s}\right\rangle ^{(T)} &=&\frac{4ea^{-D-1}}{\left( 2\pi
\right) ^{\frac{D+1}{2}}}\left( \frac{-L}{\pi }\partial _{\delta
_{0}}\right) ^{\delta _{s3}}\sideset{}{'}{\sum}_{l=0}^{\infty }\cos \left(
2\pi l\delta _{0}\right)  \notag \\
&&\times \sum_{j=1}^{\infty }\left[ 2\sideset{}{'}{\sum}%
_{k=0}^{[q/2]}A_{k}^{(s)}Z_{\nu }^{D}\left( u_{jkl}\right) -\frac{q}{\pi }%
\int_{0}^{\infty }dy\ \frac{h_{s}(q,\alpha _{0},y)Z_{\nu }^{D}\left(
u_{jyl}\right) }{\cosh (qy)-\cos (q\pi )}\right] \,,  \label{jmumu0}
\end{eqnarray}%
for $s=2,3$. Here, we have defined%
\begin{eqnarray}
u_{jkl} &=&1+\frac{j^{2}\beta ^{2}}{2w^{2}}+\frac{l^{2}L^{2}}{2w^{2}}+2\rho
^{2}s_{k}^{2},  \notag \\
u_{jyl} &=&1+\frac{j^{2}\beta ^{2}}{2w^{2}}+\frac{l^{2}L^{2}}{2w^{2}}+2\rho
^{2}\cosh ^{2}(y/2)\ .  \label{ujyl}
\end{eqnarray}%
In (\ref{jmumu0}) and in what follows $h_{3}(q,\alpha _{0},y)=h_{0}(q,\alpha
_{0},y)$ and%
\begin{eqnarray}
A_{k}^{(s)} &=&\cos \left( 2\pi k\alpha _{0}\right) ,\;s=0,3,  \label{Ak} \\
A_{k}^{(2)} &=&\sin {(2\pi k/q)}\sin \left( 2\pi k\alpha _{0}\right) . 
\notag
\end{eqnarray}

The current densities in the geometry with decompactified $z$-coordinate are
obtained from (\ref{jmumu0}) in the limit $L\rightarrow \infty $. In this
limit the axial current density vanishes and in the expression for the
azimuthal current density the contribution of the term with $l=0$ survives
only:%
\begin{equation}
\left\langle j^{2}\right\rangle _{L=\infty }^{(T)}=\frac{2ea^{-D-1}}{\left(
2\pi \right) ^{\frac{D+1}{2}}}\sum_{j=1}^{\infty }\left[ 2\sideset{}{'}{\sum}%
_{k=0}^{[q/2]}A_{k}^{(2)}Z_{\nu }^{D}\left( u_{jk0}\right) -\frac{q}{\pi }%
\int_{0}^{\infty }dy\ \frac{h_{2}(q,\alpha _{0},y)Z_{\nu }^{D}\left(
u_{jy0}\right) }{\cosh (qy)-\cos (q\pi )}\right] .  \label{jsLinf}
\end{equation}%
The expression for the total azimuthal current density in this special case
is obtained from (\ref{jsLinf}) by the replacement $\sum_{j=1}^{\infty
}\rightarrow \sideset{}{'}{\sum}_{j=0}^{\infty }$ and the part with $j=0$
will correspond to the vacuum current. As before, the prime on the sign of
the summation means that the $j=0$ term is taken with coefficient 1/2.

\subsection{Conformally coupled massless field}

Another special case when the expressions for the components of the current
density are simplified corresponds to a conformally coupled massless field.
In this case one has $\nu =1/2$ and, by taking into account that $I_{1/2}(x)=%
\sqrt{2/\pi x}\sinh x$, the current density is presented as%
\begin{equation}
\left\langle j^{s}\right\rangle ^{(T)}=\left( w/a\right) ^{D+1}\left\langle
j^{s}\right\rangle _{\mathrm{(M,b)}}^{(T)},  \label{j0cc}
\end{equation}%
where%
\begin{eqnarray}
\left\langle j^{s}\right\rangle _{\mathrm{(M,b)}}^{(T)} &=&\frac{2e}{\pi ^{%
\frac{D+1}{2}}}\int_{0}^{\infty }du\,u^{\frac{D-1}{2}}\left(
1-e^{-4w^{2}u}\right) {\mathcal{J}}_{s}(q,\alpha _{0},2ur^{2})  \notag \\
&&\times \left( -\frac{L}{\pi }\partial _{\delta _{0}}\right) ^{\delta
_{s3}}F_{1}\left( \delta _{0},L^{2}u\right) \left( 2\partial _{{\mu }%
}\right) ^{\delta _{s0}}F_{2}\left( \beta {\mu },\beta ^{2}u\right) .
\label{j0Mb}
\end{eqnarray}%
For a conformally coupled scalar field the problem under consideration is
conformally related to the problem with a cosmic string in the Minkowski
bulk described by the line element%
\begin{equation}
ds_{\mathrm{M}}^{2}=dt^{2}-dr^{2}-r^{2}d\phi
^{2}-dw^{2}-dz^{2}-\sum_{i=5}^{D}(dx^{i})^{2}\ ,  \label{ds2M}
\end{equation}%
with compactified $z$-coordinate and with an additional planar boundary at $%
w=0$ on which the scalar field obeys Dirichlet boundary condition, $\varphi
|_{w=0}=0$. The boundary $w=0$ in the Minkowskian problem is the conformal
image of the AdS boundary. The expectation value (\ref{j0Mb}) is the finite
temperature contribution to the current density in the Minkowskian problem
and (\ref{j0cc}) is the standard relation between the expectation values in
two conformally related problems. The part in (\ref{j0Mb}) coming from the
first term in $1-e^{-4w^{2}u}$ corresponds to the charge density for a
massless scalar field in the Minkowskian problem when the boundary at $w=0$
is absent (we will denote that part by $\left\langle j^{0}\right\rangle _{%
\mathrm{(M)}}^{(T)}$) and the contribution coming from the second term is
induced by the Dirichlet boundary at $w=0$.

The expression in the right-hand side of (\ref{j0Mb}) is further transformed
by using the representations (\ref{F1b}), (\ref{F2}), (\ref{Sum_I}) and (\ref%
{Kcal}). The integral over $u$ in (\ref{j0Mb}) is expressed in terms of the
modified Bessel function $K_{\nu }(x)$ and for the thermal current density
we obtain%
\begin{eqnarray}
\left\langle j^{s}\right\rangle _{\mathrm{(M,b)}}^{(T)} &=&\frac{2e}{\left(
2\pi \right) ^{\frac{D}{2}}L}\left( \frac{-L}{\pi }\partial _{\delta
_{0}}\right) ^{\delta _{s3}}\left( 2\partial _{{\mu }}\right) ^{\delta
_{s0}}\sum_{l=-\infty }^{\infty }|\tilde{k}_{l}|^{D}\sum_{j=1}^{\infty
}\cosh \left( j\beta {\mu }\right)  \notag \\
&&\times \sum_{n=0,1}(-1)^{n}\left[ 2\sideset{}{'}{\sum}%
_{k=0}^{[q/2]}A_{k}^{(s)}f_{\frac{D}{2}}\left( |\tilde{k}_{l}|\sqrt{%
j^{2}\beta ^{2}+4r^{2}s_{k}^{2}+4nw^{2}}\right) \right.  \notag \\
&&\left. -\frac{q}{\pi }\int_{0}^{\infty }dy\ \frac{f_{\frac{D}{2}}\left( |%
\tilde{k}_{l}|\sqrt{j^{2}\beta ^{2}+4r^{2}\cosh ^{2}(y/2)+4nw^{2}}\right) }{%
\cosh (qy)-\cos (q\pi )}h_{s}(q,\alpha _{0},y)\right] ,  \label{jmuMb}
\end{eqnarray}%
where%
\begin{equation}
f_{\nu }(x)=x^{-\nu }K_{\nu }(x).  \label{fnu}
\end{equation}%
In the expression for the axial current density we can use the relation%
\begin{equation}
\frac{-L}{\pi }\partial _{\delta _{0}}\left[ |\tilde{k}_{l}|^{D}f_{\frac{D}{2%
}}(|\tilde{k}_{l}|x)\right] =2\tilde{k}_{l}|\tilde{k}_{l}|^{D-2}f_{\frac{D}{2%
}-1}(|\tilde{k}_{l}|x)  \label{relf}
\end{equation}%
The $n=0$ term in (\ref{jmuMb}) presents the current density $\left\langle
j^{s}\right\rangle _{\mathrm{(M)}}^{(T)}$ and the part with $n=1$ is induced
by the boundary at $w=0$. Note that $\left\langle j^{s}\right\rangle _{%
\mathrm{(M,b)}}^{(T)}|_{w=0}=0$. This result is a consequence of Dirichlet
boundary condition imposed on the scalar field at $w=0$.

\section{Asymptotic analysis}

\label{sec:Asymp}

\subsection{Small and high temperatures}

Let us consider the behavior of the thermal currents in asymptotic regions
of the parameters. For small temperatures one has ${\beta /w\gg 1}$ and the
dominant contribution to the integral over $x$ in (\ref{jmu}) comes from the
region with small $x$. By using the asymptotic (\ref{S2as2}) for the
function $F_{1}(\delta _{0},u)$ and the asymptotic (\ref{Jcalsm}) for ${%
\mathcal{J}}_{s}(q,\alpha _{0},z)$, we get%
\begin{eqnarray}
\left\langle j^{s}\right\rangle ^{(T)} &\approx &\frac{eq(w/a)^{D+1}w^{2\nu
+1}r^{2q|\alpha _{0}|-2\delta _{s2}}\left( q\alpha _{0}\right) ^{\delta
_{s2}}T}{2^{\nu +q|\alpha _{0}|+1}\left( 2\pi \right) ^{\frac{D}{2}-1}\Gamma
\left( \nu +1\right) \Gamma (q|\alpha _{0}|+1)}  \notag \\
&&\times \left( 2\partial _{{\tilde{\mu}}}\right) ^{\delta _{s0}}\left( -%
\frac{\partial _{\delta _{0}}}{\pi }\right) ^{\delta _{s3}}\left(
E_{0}T\right) ^{\frac{D}{2}+\nu +q|\alpha _{0}|-\delta
_{s2}}\sum_{j=1}^{\infty }\frac{e^{-j\left( E_{0}-|{\mu |}\right) /T}}{j^{%
\frac{D}{2}+\nu +q|\alpha _{0}|+1-\delta _{s2}}}.  \label{jmusmallT}
\end{eqnarray}%
If in addition $T\ll E_{0}-{\mu }$, the $j=1$ term dominates and the
components of the current density tend to zero like $T^{D/2+\nu +q|\alpha
_{0}|}e^{-(E_{0}-{\mu )/T}}$.

At large temperatures, $wT\gg 1$, the discussion of the asymptotic behavior
differs for the charge density and currents. That is related to the
different behavior of the integrand at the upper limit of integration in (%
\ref{jmu}). For the charge density, we introduce in (\ref{j03}) a new
variable $x=\beta ^{2}u$ and expand the functions with large arguments $%
x/\beta ^{2}$. By using (\ref{S2as}) for $F_{1}(\delta _{0},u)$, the leading
term is expressed as%
\begin{equation}
\left\langle j^{0}\right\rangle ^{(T)}\approx \left( \frac{w}{a}\right)
^{D+1}\frac{2e{\mu }}{\pi ^{\frac{D+1}{2}}}\Gamma \left( \frac{D+1}{2}%
\right) \zeta \left( D-1\right) {T}^{D-1}.  \label{j0largeT}
\end{equation}%
For the components $s=2,3$ we use the asymptotic (\ref{S2as2}) for the
function $F_{2}\left( \beta {\mu },\beta ^{2}u\right) $. To the leading
order this gives 
\begin{eqnarray}
\left\langle j^{s}\right\rangle ^{(T)} &\approx &\frac{2eTw}{\pi ^{\frac{D-1%
}{2}}}\left( \frac{w}{a}\right) ^{D+1}\int_{0}^{\infty }du\,u^{\frac{D-1}{2}}%
\frac{I_{\nu }(2w^{2}u)}{e^{2w^{2}u}}\exp \left( \frac{{\mu }^{2}}{4u}\right)
\notag \\
&&\times {\mathcal{J}}_{s}(q,\alpha _{0},2ur^{2})\left( \frac{-L}{\pi }%
\partial _{\delta _{0}}\right) ^{\delta _{s3}}F_{1}\left( \delta
_{0},L^{2}u\right) .  \label{jmulargeT}
\end{eqnarray}

It is of interest to compare the obtained results with the thermal charge
density in the Minkowski spacetime. For a scalar field with mass $m$ and the
chemical potential ${\mu }$ the expectation value of the charge density is
given by (see, for example, \cite{Beze13T}) 
\begin{equation}
\left\langle j^{0}\right\rangle _{\mathrm{(M)}}^{(T)}=\frac{4e\beta m^{D+1}}{%
(2\pi )^{\frac{D+1}{2}}}\sum_{j=1}^{\infty }j\sinh (j{\mu }\beta )f_{\frac{%
D+1}{2}}(j\beta m),  \label{j0M}
\end{equation}%
where $f_{\nu }(x)$ is defined by (\ref{fnu}). At high temperatures $T\gg m$
for the leading order term we get 
\begin{equation}
\left\langle j^{0}\right\rangle _{\mathrm{(M)}}^{(T)}\approx \frac{2e{\mu }}{%
\pi ^{\frac{D+1}{2}}}\Gamma \left( \frac{D+1}{2}\right) \zeta \left(
D-1\right) T^{D-1}.  \label{j0MhighT}
\end{equation}%
Comparing with (\ref{j0largeT}), we see the relation $\left\langle
j^{0}\right\rangle ^{(T)}\approx (w/a)^{D+1}\left\langle j^{0}\right\rangle
_{\mathrm{(M)}}^{(T)}$ at high temperatures. At high temperatures the
dominant contribution to the charge density comes from the field
fluctuations with small wavelengths and the effects of the curvature are
small.

\begin{figure}[tbph]
\begin{center}
\epsfig{figure=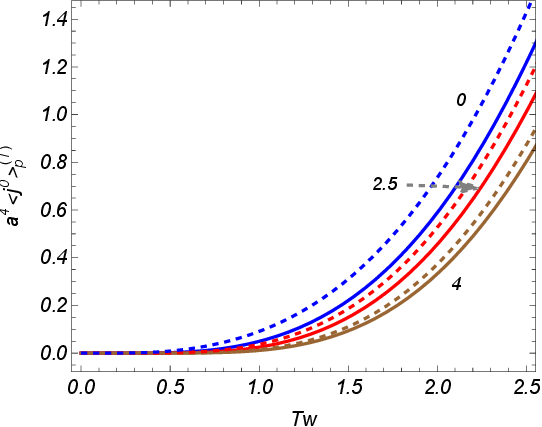,width=7.5cm,height=6cm}
\end{center}
\caption{The charge density for minimally (full curves) and conformally
(dashed curves) coupled scalar fields in the model with $D=4$ versus the
temperature for fixed values $r/w=0.5$, $L/w=1$, $w\protect\mu =0.5$, $q=1.5$%
, $\protect\alpha _{0}=\protect\delta _{0}=0.2$. The graphs are plotted for
different values of the product $ma$.}
\label{fig2}
\end{figure}

\begin{figure}[tbph]
\begin{center}
\begin{tabular}{cc}
\epsfig{figure=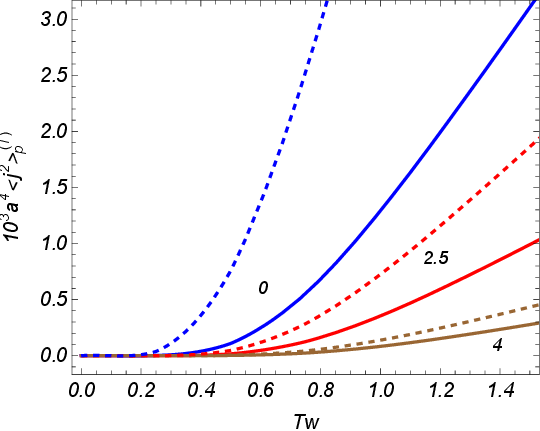,width=7.5cm,height=6cm} & \quad %
\epsfig{figure=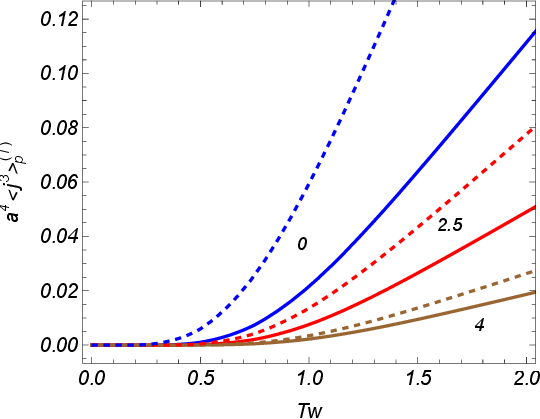,width=7.5cm,height=6cm}%
\end{tabular}%
\end{center}
\caption{The same as in figure \protect\ref{fig2} for the azimuthal (left
panel) and axial (right panel) current densities.}
\label{fig3}
\end{figure}

Figures \ref{fig2} and \ref{fig3} display the thermal charge and current
densities as functions of the temperature in the $D=4$ model for $r/w=0.5$, $%
L/w=1$, $w\mu =0.5$, $q=1.5$, $\alpha _{0}=\delta _{0}=0.2$. The full and
dashed curves correspond to minimally and conformally coupled fields and the
numbers near the curves are the values of $ma$. The numerical data presented
in figures confirm the features clarified by the asymptotic analysis. Liner
dependence for the azimuthal and axial current densities and stronger
increase of the charge density at high temperatures are clearly seen.

\subsection{Asymptotics with respect to the compactification length}

For small values of the compactification length, $L/w\ll 1$, we use the
asymptotic (\ref{S2as2}) for the function $F_{1}(\delta _{0},u)$. The
dominant contribution to the $u$-integral in (\ref{jmu}) comes from the
region $u\gtrsim \pi ^{2}|\delta _{0}|^{2}/L^{2}$. In that region $u$ is
large and we use the corresponding asymptotic for the integrand. The
integral over $u$ is expressed in terms of the modified Bessel function and
we get%
\begin{eqnarray}
\left\langle j^{s}\right\rangle ^{(T)} &\approx &\frac{2eE_{0}^{D}}{\left(
2\pi \right) ^{\frac{D}{2}}L}\left( \frac{w}{a}\right) ^{D+1}\left( \frac{-L%
}{\pi }\partial _{\delta _{0}}\right) ^{\delta _{s3}}\left( 2\partial _{{%
\tilde{\mu}}}\right) ^{\delta _{s0}}\cosh \left( \beta {\mu }\right)  \notag
\\
&&\left[ 2\sideset{}{'}{\sum}_{k=0}^{[q/2]}A_{k}^{(s)}f_{\frac{D}{2}}(E_{0}%
\sqrt{\beta ^{2}+4r^{2}s_{k}^{2}})\right.  \notag \\
&&\left. -\frac{q}{\pi }\int_{0}^{\infty }dy\ \frac{f_{\frac{D}{2}}(E_{0}%
\sqrt{\beta ^{2}+4r^{2}\cosh ^{2}(y/2)})}{\cosh (qy)-\cos (q\pi )}%
h_{s}(q,\alpha _{0},y)\right] .  \label{jmusmallL}
\end{eqnarray}%
If in addition, $\beta /L,r/L\gg 1$, for the components with $s=0,3$ the
main contribution in (\ref{jmusmallL}) comes from the term $k=0$ and they
behave as $\left\langle j^{s}\right\rangle ^{(T)}\propto e^{-E_{0}\beta }/L^{%
\frac{D+1}{2}}$. For the component $\left\langle j^{2}\right\rangle ^{(T)}$
one has $A_{0}^{(2)}=0$ and an additional exponential suppression factor is
present coming from the part in the square brackets of (\ref{jmusmallL}).
Hence, for $\delta _{0}>0$ and for small $L$ we have an exponential
suppression for all the components.

For $\delta _{0}=0$ we should also put $\mu =0$. In this case the only
nonzero component corresponds to the azimuthal current density. In order to
find the asymptotic for $L/w\ll 1$ we use the expression (\ref{jmumu0}) for $%
s=2$ and $\delta _{0}=0$. The dominant contribution to the series over $l$
comes from large values of $l$ and to the leading order we can replace the
summation by the integration: $\sum_{l=0}^{\infty }\rightarrow
\int_{0}^{\infty }dl$. The resulting integral is evaluated by using the
formula%
\begin{equation}
\int_{0}^{\infty }du\,Z_{\nu }^{D}\left( b^{2}+u^{2}\right) =\frac{\sqrt{\pi 
}}{2}Z_{\nu }^{D-1}(b^{2}).  \label{IntZ1}
\end{equation}
This formula is obtained by using the integral representation for the
function $Z_{\nu }^{D}\left( x\right) $ given by (\ref{IntZ}). In this way
we can see that%
\begin{equation}
\left\langle j^{2}\right\rangle ^{(T)}|_{\delta _{0}=0}\approx \frac{w}{aL}%
\left\langle j^{2}\right\rangle _{D}^{(T)},  \label{j2smallL}
\end{equation}%
where $\left\langle j^{2}\right\rangle _{D}^{(T)}$ is the azimuthal current
density for a scalar field with zero chemical potential in a $D$-dimensional
AdS spacetime, obtained from the geometry described above excluding the $z$%
-coordinate. The corresponding expression is obtained from (\ref{jsLinf}) by
the replacement $D\rightarrow D-1$.

Considering the asymptotic for large values of the compactification length,
we note that for a given value of the chemical potential the maximal length
of the compact dimension, $L_{m}$, is determined by the condition (\ref%
{mucond}): $L_{m}=2\pi |\delta _{0}|/|\mu |$. For the zero chemical
potential, ${\mu }{=0}$, the thermal charge density vanishes. Denoting $%
\left\langle j^{s}\right\rangle _{L=\infty }^{(T)}=\lim_{L\rightarrow \infty
}\left\langle j^{s}\right\rangle ^{(T)}$ for the components $s=2,3$, we can
see that $\left\langle j^{3}\right\rangle _{L=\infty }^{(T)}=0$ and 
\begin{equation}
\left\langle j^{2}\right\rangle _{L=\infty }^{(T)}=\frac{ea^{-D-1}}{\left(
2\pi \right) ^{\frac{D}{2}}}\int_{0}^{\infty }dx\,x^{\frac{D}{2}}\frac{%
I_{\nu }(x)}{e^{x}}{\mathcal{J}}_{2}(q,\alpha _{0},x\rho ^{2})F_{2}\left(
0,\beta ^{2}x/(2w^{2}\right) ).  \label{j2Linf}
\end{equation}%
This result gives the current density for a cosmic string in locally AdS
bulk where the $z$-direction has uncompactified topology $R^{1}$. In order
to find the next terms in the expansion over $1/L$, we use the
representation (\ref{jmu}) with the function $F_{1}\left( \delta
_{0},L^{2}u\right) $ from (\ref{F1}) and with 
\begin{equation}
F_{2}\left( 0,\beta ^{2}u\right) =\frac{1}{\beta }\sqrt{\frac{\pi }{u}}%
\sideset{}{'}{\sum}_{j=0}^{\infty }e^{-\pi ^{2}j^{2}/(\beta ^{2}u)}-\frac{1}{%
2}.  \label{F20}
\end{equation}%
The leading contribution to the difference $\left\langle j^{s}\right\rangle
^{(T)}-\left\langle j^{s}\right\rangle _{L=\infty }^{(T)}$ comes from the
term $j=0$ in (\ref{F20}) and we get%
\begin{eqnarray}
\left\langle j^{s}\right\rangle ^{(T)} &\approx &\left\langle
j^{s}\right\rangle _{L=\infty }^{(T)}+\frac{2eqw^{2\nu +1}r^{2q|\alpha
_{0}|}\left( w/a\right) ^{D+1}T}{\pi ^{\frac{D-1}{2}}\Gamma \left( \nu
+1\right) \Gamma (q|\alpha _{0}|+1)}\left( \frac{q\alpha _{0}}{2r^{2}}%
\right) ^{\delta _{s2}}  \notag \\
&&\times \left( \frac{-L}{\pi }\partial _{\delta _{0}}\right) ^{\delta
_{s3}}\sum_{l=1}^{\infty }\cos \left( 2\pi l\delta _{0}\right) \frac{\Gamma
\left( \frac{D+1}{2}+\nu +q|\alpha _{0}|-\delta _{s2}\right) }{\left(
lL\right) ^{D+1+2\left( \nu +q|\alpha _{0}|-\delta _{s2}\right) }}.
\label{jmuLlarge}
\end{eqnarray}%
for $s=2,3$.

\begin{figure}[tbph]
\begin{center}
\epsfig{figure=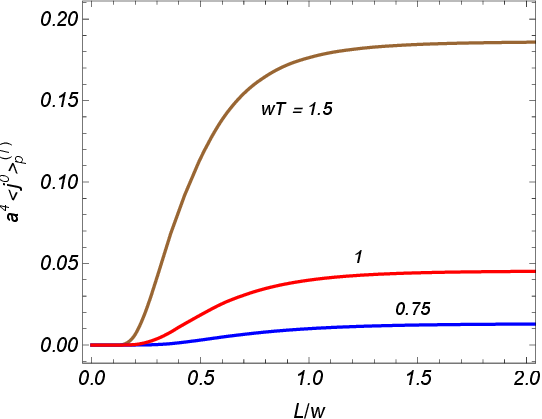,width=7.5cm,height=6cm}
\end{center}
\caption{The charge density for a minimally coupled scalar field as a
function of the compactification length in the $D=4$ model. The graphs are
plotted for $r/w=0.5$, $w\protect\mu =0.5$, $q=1.5$, $\protect\alpha %
_{0}=0.25$, $\protect\delta _{0}=1/3$ and the numbers near the curves are
the respective values of the temperature. }
\label{fig4}
\end{figure}

\begin{figure}[tbph]
\begin{center}
\begin{tabular}{cc}
\epsfig{figure=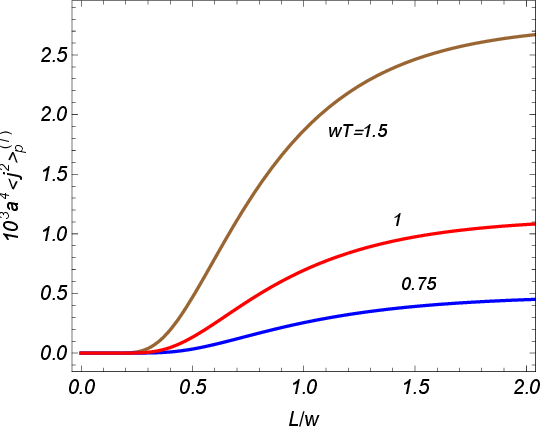,width=7.5cm,height=6cm} & \quad %
\epsfig{figure=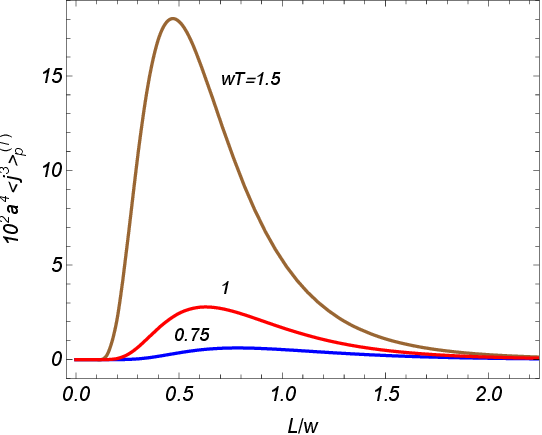,width=7.5cm,height=6cm}%
\end{tabular}%
\end{center}
\caption{The same as in figure \protect\ref{fig4} for the azimuthal (left
panel) and axial (right panel) current densities.}
\label{fig5}
\end{figure}

For $D=4$ minimally coupled massless scalar field, the dependence of the
charge and current densities on the proper length of the compact dimension
is depicted in Figures \ref{fig4} and \ref{fig5}. The graphs are plotted for 
$r/w=0.5$, $w\mu =0.5$, $q=1.5$, $\alpha _{0}=0.25$, $\delta _{0}=1/3$ and
for different values of the temperature (the numbers near the curves). In
accordance with the asymptotic analysis, all the components tend to zero for
small values of the compactification length.

\subsection{Small and large distances from cosmic string}

In order to find the asymptotic of the thermal current density near the
cosmic string, $\rho \ll 1$, we use the expression (\ref{jmu}) and the
approximation (\ref{Jcalsm}) for the function ${\mathcal{J}}_{s}(q,\alpha
_{0},2ur^{2})$. The latter shows that $\left\langle j^{s}\right\rangle
^{(T)}\propto r^{2q|\alpha _{0}|-2\delta _{s2}}$ for $r\rightarrow 0$.
Hence, for $\alpha _{0}\neq 0$ the charge and axial current densities vanish
on the string. The contravariant component of the azimuthal current density
vanishes on the string for $|\alpha _{0}|>1/q$ and diverges for $|\alpha
_{0}|<1/q$. Note that the physical component of the azimuthal current
density behaves as $\left\langle j^{2}\right\rangle _{\mathrm{p}%
}^{(T)}\propto r^{2q|\alpha _{0}|-1}$.

At large distances from the string, $\rho \gg 1$, one gets $%
\lim_{r\rightarrow \infty }\left\langle j^{s}\right\rangle
^{(T)}=\left\langle j^{s}\right\rangle _{0}^{(T)}$, where the components
with $s=0,3$ are given by (\ref{jmu10}) and $\left\langle j^{2}\right\rangle
_{0}^{(T)}=0$. The effects induced by cosmic string and magnetic flux are
encoded in the difference $\left\langle j^{s}\right\rangle
^{(T)}-\left\langle j^{s}\right\rangle _{0}^{(T)}$. For $\rho \gg 1$ the
dominant contribution to the integral over $u$ in (\ref{jmu}) comes from the
region with small values of $u$. By using the corresponding asymptotics (\ref%
{S2as2}), we can see that%
\begin{eqnarray}
\left\langle j^{s}\right\rangle ^{(T)} &=&\left\langle j^{s}\right\rangle
_{0}^{(T)}+\frac{2eTw^{D+2\nu +2}{\mu }^{\delta _{s0}}}{\pi ^{\frac{D}{2}%
-1}\Gamma \left( \nu +1\right) a^{D+1}L}\left( \frac{2\pi }{L}\delta
_{0}\right) ^{\delta _{s3}}\left( \frac{E_{0}^{2}-{\mu }^{2}}{2}\right) ^{%
\frac{D}{2}+\nu -1+\delta _{s2}}  \notag \\
&&\times \left[ 2\sideset{}{'}{\sum}_{k=1}^{[q/2]}A_{k}^{(s)}f_{\frac{D}{2}%
+\nu -1+\delta _{s2}}\left( 2rs_{k}\sqrt{E_{0}^{2}-{\mu }^{2}}\right) \right.
\notag \\
&&\left. -\frac{q}{\pi }\int_{0}^{\infty }dy\ \frac{f_{\frac{D}{2}+\nu
-1+\delta _{s2}}\left( 2r\sqrt{E_{0}^{2}-{\mu }^{2}}\cosh (y/2)\right) }{%
\cosh (qy)-\cos (q\pi )}h_{s}(q,\alpha _{0},y)\right] ,  \label{jmularger}
\end{eqnarray}%
If in addition $r\sqrt{E_{0}^{2}-{\mu }^{2}}\gg 1$, the cosmic string
induced contribution $\left\langle j^{s}\right\rangle ^{(T)}-\left\langle
j^{s}\right\rangle _{0}^{(T)}$ decays as $e^{-2rs_{1}\sqrt{E_{0}^{2}-{\mu }%
^{2}}}/r^{(D-1)/2+\nu +\delta _{s2}}$ for $q\geq 2$ and like $e^{-2r\sqrt{%
E_{0}^{2}-{\mu }^{2}}}/r^{(D-1)/2+\nu +\delta _{s2}}$ for $1<q<2$. As it has
been already discussed in Section \ref{Current}, the total charge, per unit
volume in the subspace $(\mathbf{x}_{\parallel },w)$, induced by the
presence of the cosmic string is finite and is given by the expression (\ref%
{Qcsd}). This finiteness is a consequence of the exponential decrease of the
related charge density at large distances from the string.

\begin{figure}[tbph]
\begin{center}
\epsfig{figure=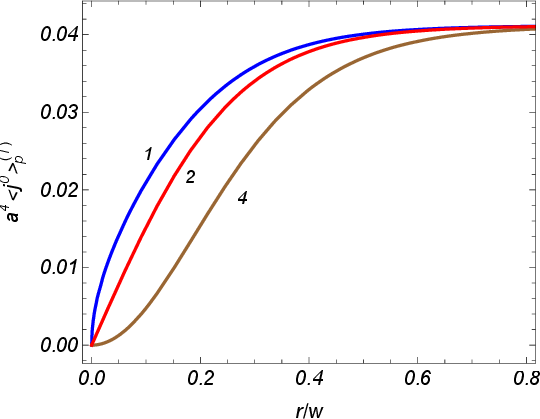,width=7.5cm,height=6cm}
\end{center}
\caption{The dependence of the thermal charge density on the proper distance
from the cosmic string for a minimally coupled $D=4$ scalar field. For the
parameters we have taken $L/w=1$, $Tw=1$, $w\protect\mu =0.5$, $\protect%
\alpha _{0}=0.25$, $\protect\delta _{0}=1/3$ and the numbers near the curves
present the values of $q$. }
\label{fig6}
\end{figure}

\begin{figure}[tbph]
\begin{center}
\begin{tabular}{cc}
\epsfig{figure=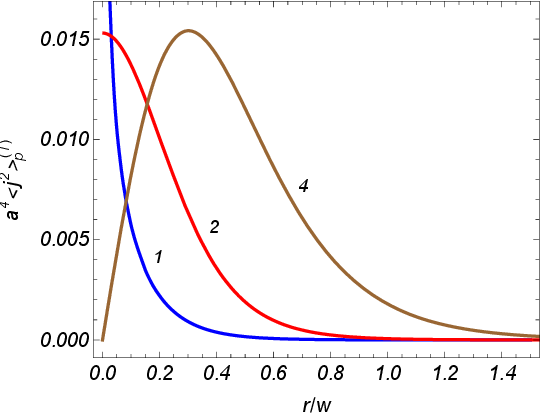,width=7.5cm,height=6cm} & \quad %
\epsfig{figure=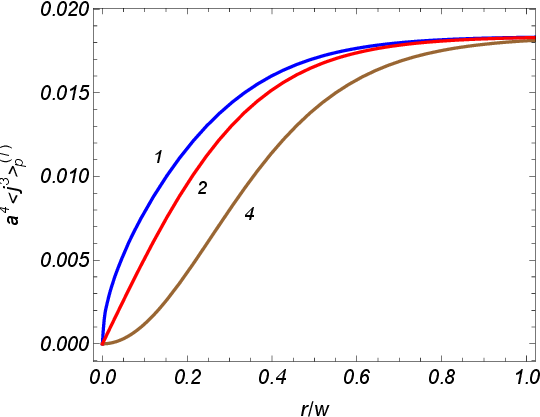,width=7.5cm,height=6cm}%
\end{tabular}%
\end{center}
\caption{The same as in figure \protect\ref{fig6} for the azimuthal (left
panel) and axial (right panel) current densities.}
\label{fig7}
\end{figure}

The dependence of the thermal charge and current densities on the proper
distance from the cosmic string is displayed in Figures \ref{fig6} and \ref%
{fig7}. The graphs are plotted for $D=4$ minimally coupled massless field
for fixed values $L/w=1$, $Tw=1$, $w\mu =0.5$, $\alpha _{0}=0.25$, $\delta
_{0}=1/3$ and the numbers near the curves correspond to the values of the
parameter $q$. In accordance with the asymptotic analysis given above, the
components with $s=0,3$ vanish on the string. For the parameters we have
taken in Figure \ref{fig7} and for $q=2$ one has $2q|\alpha _{0}|=1$ and the
azimuthal component tends to finite nonzero value on the cosmic string. The
graphs on the left panel of Figure \ref{fig7} also confirm the features
described by the asymptotic analysis on that the physical component of the
thermal contribution to the azimuthal current density vanishes on the string
for $|\alpha _{0}|>1/(2q)$ and diverges for $|\alpha _{0}|<1/(2q)$.

Before we summarize the main results of the paper it is worth pointing out
about some potential applications and generalizations. In Introduction we
have mentioned two mechanisms for the production of cosmic string type
topological defects: formation of cosmic strings as a result of symmetry
breaking phase transitions in the expanding early universe and formation of
string-type linear structures in brane inflation models. The defects
produced in the second mechanism can be either fundamental strings or
one-dimensional branes (F- and D-strings, respectively). Their bound states
can be formed as well. The linear structures formed in brane inflation
models may have different sizes and cosmologically extended ones become
cosmic superstrings. The investigations of the effects induced by those
structures in the AdS spacetime are partly motivated by that the related
geometry appears as ground state in superstring theories and as a bulk
geometry in braneworld models. An additional motivation comes from the
AdS/CFT correspondence: the initial version of that correspondence states a
duality between IIB string theory on the AdS bulk and $\mathcal{N}=4$
super-Yang-Mills (SYM) theory on its boundary. The defect we have discussed
in the present paper can be considered as a simple realization of D$p$-brane
with $p=D-2$. The geometry of the core is described by the line element (\ref%
{dsc}) that corresponds to $(D-1)$-dimensional AdS spacetime (locally AdS in
the model with compactified $z$-direction). The vacuum polarization and the
Casimir forces for planar D$p$-branes with $p=D-1$, orthogonal to the AdS
boundary, have been discussed in Refs. \cite{Beze15Ort,Bell22Ort}. In the
problem under consideration and for $D=3$ one has a linear defect in the AdS
bulk and a point-like defect in the dual theory on the AdS boundary. Note
that the gravitational field of the point mass in 3-dimensional spacetime is
described by conical geometry (see, for example, \cite{Dese84}). That
geometry also appears in the effective long-wavelength description for a
number of planar condensed matter systems (examples are the graphene
nanocones). In Ref. \cite{Bayo11} the AdS$_{4}$/CFT$_{3}$ correspondence
have been used to evaluate the Green function for scalar operators in
3-dimensional conical spacetime. For spatial dimension $D=4$ the defect in
the bulk is 2-dimensional and the corresponding counterpart in the dual
theory is the standard straight cosmic string (in the model with $-\infty
<z<+\infty $) or a cosmic string compactified along its axis (model with
compactified $z$-direction). Hence, the results described in the present
paper can be used to investigate the thermal effects of cosmic string type
defects in the dual theory on the AdS boundary by using the AdS/CFT map.

We have considered a thermal field theory in the background geometry
described by the metric (\ref{ds1}). In the context of the AdS/CFT
correspondence, the dual theory is a thermal field theory with the line
element $ds_{\mathrm{CFT}}^{2}=dt^{2}-\gamma _{ij}dx^{i}dx^{j}$, where $%
\gamma _{ij}dx^{i}dx^{j}=dr^{2}+r^{2}d\phi ^{2}+dz^{2}+d\mathbf{x}%
_{\parallel }^{2}$. Note that for the geometry with decompactified $z$%
-coordinate and for scale invariant dual theories the only dimensionful
parameter is the temperature. The latter can be changed by a scaling and
there are no phase transitions in those theories. An example is the $%
\mathcal{N}=4$ SYM on $R^{3}$ as the dual theory in the initial version of
the AdS/CFT correspondence. The compactification of the $z$-coordinate
introduces an additional length scale in the model and the phase transitions
may take place (see, for example, the discussion in Ref. \cite{Nats15} for $%
\mathcal{N}=4$ SYM). It is important to note that the AdS counterpart of the
boundary theory is not unique. As alternatives to the pure AdS spacetime,
AdS black holes with different geometries of the horizon have been
considered in the literature (for black hole solutions in AdS spacetime with
different topologies of the horizon see, e.g., Refs. \cite{Mann97,Birm99}).
The corresponding solution with the planar horizon, most appropriate in the
problem under consideration, is given by%
\begin{equation}
ds_{\mathrm{BH}}^{2}=\frac{a^{2}}{w^{2}}\left\{ \left[ 1-\left( \frac{w}{w_{%
\mathrm{H}}}\right) ^{D}\right] dt^{2}-\frac{dw^{2}}{1-\left( w/w_{\mathrm{H}%
}\right) ^{D}}-\gamma _{ij}dx^{i}dx^{j}\right\} ,  \label{ds2BH}
\end{equation}%
where the hypersurface $w=w_{\mathrm{H}}$, with 
\begin{equation}
w_{\mathrm{H}}=\left[ \frac{(D-1)a^{D-1}}{16\pi G\rho _{m}}\right] ^{1/D},
\label{wH}
\end{equation}%
corresponds to the horizon. Here, $G$ is the gravitational constant in $%
(D+1) $-dimensional spacetime and $\rho _{m}$ is the mass per unit volume of
the subspace with the line element $\gamma _{ij}dx^{i}dx^{j}$. For $0\leq
\phi \leq 2\pi $, introducing a new radial coordinate $\rho =a^{2}/w$ and
the Cartesian coordinates in the subspace $(r,\phi )$, this solution is
reduced to the one given, for example, in Ref. \cite{Birm99}. In the case $%
0\leq \phi \leq \phi _{0}<2\pi $ the metric (\ref{ds2BH}) describes a
combination of AdS planar black hole with a generalized cosmic string type
topological defect crossing the black hole horizon. In the context of the
AdS/CFT correspondence, two geometries described by (\ref{ds2}) and (\ref%
{ds2BH}) may correspond to two different phases of the dual theory. The
phase transition in the dual theory is interpreted in terms of the
Hawking-Page phase transition in the AdS bulk \cite{Hawk83}.

The thermal charge and current densities will also appear for a quantum
charged scalar field propagating in the background of the black hole
geometry (\ref{ds2BH}). At large distances from the horizon, $w/w_{\mathrm{H}%
}\ll 1$, corresponding to points near the AdS boundary, the geometry we have
considered in the present paper is the leading order approximation to the
line element (\ref{ds2BH}). The expressions for the charge and current
densities given above will give the leading terms in the respective
asymptotic expansions with respect to the small ratio $w/w_{\mathrm{H}}$.
Another limiting region where the geometry (\ref{ds2BH}) is simplified
corresponds to points near the black hole horizon, $1-w/w_{\mathrm{H}}\ll 1$%
. Expanding the metric tensor and introducing new coordinates%
\begin{equation}
\tau =\frac{Dt}{2w_{\mathrm{H}}},\;\bar{u}=-\frac{2a}{\sqrt{D}}\sqrt{1-\frac{%
w}{w_{\mathrm{H}}}},\;\left( \bar{r},\bar{z},\mathbf{\bar{x}}_{\parallel
}\right) =\frac{a}{w_{\mathrm{H}}}\left( r,z,\mathbf{x}_{\parallel }\right) ,
\label{Rind}
\end{equation}%
the line element (\ref{ds2BH}) is approximated by%
\begin{equation}
ds_{\mathrm{BH}}^{2}\approx \bar{u}^{2}d\tau ^{2}-d\bar{u}^{2}-d\bar{r}^{2}-%
\bar{r}^{2}d\phi ^{2}-d\bar{z}^{2}-d\mathbf{\bar{x}}_{\parallel }^{2}.
\label{Rind2}
\end{equation}%
The right-hand side corresponds to the $(D+1)$-dimensional Rindler spacetime
with compact $z$-direction and in the presence of a generalized cosmic
string. Note that the length of the compact dimension $\bar{z}$ is given by $%
\bar{L}=aL/w_{\mathrm{H}}$ and it depends on the location of the horizon.
The simple form of the near-horizon metric allows to obtain closed analytic
expressions for the expectation values of the charge and current densities.
In the absence of the cosmic string, the corresponding vacuum expectation
values in Rindler spacetime with an arbitrary number of toroidally compact
dimensions have been recently discussed in \cite{Kota22}.

\section{Conclusions}

\label{sec:Conc}

We have investigated combined effects of background gravitational field,
nontrivial spatial topology and finite temperature on the expectation values
of the charge and current densities for a massive scalar field with general
curvature coupling. Motivated by high symmetry and by importance in recent
theoretical constructions, as a bulk geometry we have chosen the locally AdS
spacetime. Two sources of nontrivial topology are considered. The first one
corresponds to a defect that is a generalization of a cosmic string in the
presence of extra dimensions and the second one comes from the
compactification of a spatial dimension in Poincar\'{e} coordinates. We have
started the investigation from the Hadamard two-point function. The finite
temperature part of that function in the problem at hand is expressed as (%
\ref{wight_T}) with particle and antiparticle contributions coming from the
terms with $u=+$ and $u=-$. The evaluation of various physical
characteristics bilinear in the field is based on the Hadamard function. An
important difference from the Minkowskian bulk is related to the upper bound
of the chemical potential given by (\ref{mucond}). It does not depend on the
field mass and becomes zero in the decompactification limit. This
qualitative difference is related to that the mass does not enter in the
expression for the energy of the modes. We have also evaluated the
bulk-to-boundary propagator for a scalar field that plays an important role
in the interpretations of the bulk quantities in terms of the boundary field
theory in the context of the AdS/CFT correspondence.

As important characteristics for a charged field we have considered the
charge and current densities. The vacuum expectation values of those
quantities have been considered in a recent publication \cite{Oliv19} and
here we were mainly concerned with the thermal contributions. The nonzero
components correspond to the charge density, azimuthal current density and
the current density along the compact dimension (axial current). We note
that the nonzero charge density is a pure thermal effect: the corresponding
vacuum average is zero. The combined expression for the nonzero components
is given by (\ref{jmu}). As a consequence of the AdS maximal symmetry, the
thermal expectation values depend on the parameters having the dimension of
length through the ratios like $r/w$, $L/w$. They correspond to the proper
distance from the defect and proper length of the compact dimension in the
units of the curvature radius. The charge density is an odd function of the
chemical potential. It vanishes for zero chemical potential as a result of
cancellation between the contributions from particles and antiparticles. The
azimuthal and axial currents are even functions of the chemical potential
and survive in the limit $\mu =0$. We have considered two types of magnetic
fluxes. The first one is confined inside the defect core and another one is
interpreted as a flux enclosed by compact dimension. All the components of
the current density are periodic functions of those fluxes with the flux
quantum being the period. The dependence of the expectation values on the
magnetic fluxes appears in the form of two parameters $\alpha _{0}$ and $%
\delta _{0}$ corresponding to the ratios of the magnetic fluxes to the flux
quantum.

Our general results include various special cases. In the absence of the
cosmic string the only source of the nontrivial topology is the
compactification of the $z$ coordinate and the azimuthal current density
becomes zero. The thermal contributions to the charge and axial current
densities in this special case are given by (\ref{jmu10}). For the zero
chemical potential we have provided alternative representation (\ref{jmumu0}%
) for the azimuthal and axial currents. In this special case we can consider
also the decompactification limit $L\rightarrow \infty $. For the nonzero
chemical potential one has the maximal value of the compactification length
given by $L_{m}=2\pi |\delta _{0}|/|\mu |$. For a conformally coupled
massless scalar field the problem at hand is conformally related to the
problem of a cosmic string in the Minkowski spacetime with a compact
dimension in the presence of a planar Dirichlet boundary orthogonal to the
string (see (\ref{j0cc}) and (\ref{j0Mb})). The total charge, per unit
volume in the subspace $(\mathbf{x}_{\parallel },w)$, induced by the cosmic
string is finite. Its dependence on the parameters of the cosmic string is
factorized in a simple form (see (\ref{Qcsd})). Depending on the planar
angle deficit and magnetic flux, the presence of the cosmic string can
either increase or decrease the total charge.

To clarify the behavior of the expectation values we have considered various
asymptotic limits. In the zero temperature limit the thermal densities tend
to zero like $T^{D/2+\nu +q|\alpha _{0}|}e^{-(E_{0}-{\mu )/T}}$. The high
temperature behavior is different for the charge density and current
densities. At high temperatures the influence of the gravitational field and
topology on the charge density is subdominant and the leading term coincides
with that for the charge density in the Minkowski spacetime with the
behavior $\propto T^{D-1}$. The current densities are topology-induced
quantities and their behavior at high temperatures is completely different
with the linear dependence on the temperature. For $\delta _{0}>0$ all the
expectation values are exponentially small for small values of the
compactification length. The situation is different in the case $\delta
_{0}=0$ when the chemical potential is zero as well. The charge and axial
current densities are zero in this case. The leading term in the expansion
of the azimuthal current density for small values of $L$ is given by the
right-hand side of (\ref{j2smallL}), where $\left\langle j^{2}\right\rangle
_{D}^{(T)}$ is the azimuthal current density in $D$-dimensional AdS
spacetime which is obtained from the geometry we consider excluding the
coordinate corresponding to the compact dimension $z$. In the consideration
of the decompactification limit for the $z$ direction the chemical potential
should be set zero. In the decompactified geometry the charge and axial
current densities are zero and the azimuthal current density is given by (%
\ref{j2Linf}). The next-to-leading terms in the azimuthal ($s=2$) and axial (%
$s=3$) current densities decay like $1/L^{D+1+2(\nu +q|\alpha _{0}|-\delta
_{s2})}$. The suppression for the axial component is stronger. Near the
cosmic string the thermal contributions to the charge and axial current
densities behave like $r^{2q|\alpha _{0}|}$ and they vanish on the cosmic
string for $\alpha _{0}\neq 0$. In the same limit the physical azimuthal
component behaves as $r^{2q|\alpha _{0}|-1}$. The leading terms in the
expansion of the thermal expectation values over the inverse distance from
the cosmic string coincide with the corresponding quantities in the geometry
where the cosmic string is absent. The next-to-leading terms encode the
effects of the cosmic string. For $q\geq 2$, at large distances their decay
is described by the factor $e^{-2rs_{1}\sqrt{E_{0}^{2}-{\mu }^{2}}%
}/r^{(D-1)/2+\nu +\delta _{s2}}$. The suppression factor in the range $1<q<2$
is obtained by the replacement $s_{1}\rightarrow 1$.

The investigation of thermal effects on the AdS bulk may shed light on the
phase structure in the boundary theory. We have emphasized the importance of
compactification to have phase transitions in scale invariant theories. The
confining phase in the boundary gauge theory is mapped onto AdS black hole
geometry. Our results approximate the corresponding expectation values
around black holes with planar horizons at large distances from the horizon.
We have noted that closed analytic expressions can also be obtained for the
near-horizon region where the geometry is approximated by Rindler spacetime with a compact dimension in the presence of cosmic string type topological defect.

\section*{Acknowledgment}

E.R.B.M is partially supported by CNPq under Grant no 301.783/2019-3. A.A.S.
was supported by the grants No. 20RF-059 and No. 21AG-1C047 of the Science
Committee of the Ministry of Education, Science, Culture and Sport RA.

\end{document}